\documentclass[11pt,a4paper]{article}
\pdfoutput=1

\usepackage{jheppub}
\usepackage[utf8]{inputenc}
\usepackage[T1]{fontenc}

\newcommand{\Nuc}[2]{\ensuremath{\mbox{}^{#1}\text{#2}}}
\newcommand{\Dmq}{\Delta m^2}
\newcommand{\Eps}{\varepsilon}
\newcommand{\dd}{\mathrm{d}}

\renewcommand{\Re}{\mathop{\mathrm{Re}}}
\renewcommand{\Im}{\mathop{\mathrm{Im}}}
\DeclareMathOperator{\Evol}{Evol}
\DeclareMathOperator{\Tr}{Tr}
\DeclareMathOperator{\diag}{diag}

\makeatletter
\DeclareRobustCommand\recite[1]{\begingroup\@fileswfalse\cite{#1}\endgroup}
\gdef\@fpheader{} 
\makeatother

\graphicspath{{Figures/}}

\allowdisplaybreaks

\title{Constraining New Physics with Borexino Phase-II spectral data}

\author[a]{Pilar Coloma,}
\affiliation[a]{Instituto de F\'isica Te\'orica (IFT-CFTMAT),
  CSIC-UAM, Calle de Nicol\'as Cabrera 13--15, Campus de Cantoblanco,
  E-28049 Madrid, Spain}
\emailAdd{pilar.coloma@ift.csic.es}

\author[b,c,d]{M.~C.~Gonzalez-Garcia,}
\affiliation[b]{C.N.~Yang Institute for Theoretical Physics, Stony
  Brook University, Stony Brook, NY 11794-3840, USA}
\affiliation[c]{Departament de F\'isica Qu\`antica i Astrof\'isica and
  Institut de Ci\`encies del Cosmos, Universitat de Barcelona,
  Diagonal 647, E-08028 Barcelona, Spain}
\affiliation[d]{Instituci\'o Catalana de Recerca i Estudis Avançats
  (ICREA), Pg.\ Lluis Companys 23, E-08010 Barcelona, Spain}
\emailAdd{maria.gonzalez-garcia@stonybrook.edu}

\author[a]{Michele Maltoni,}
\emailAdd{michele.maltoni@csic.es}

\author[c]{Jo\~ao Paulo Pinheiro,}
\emailAdd{joaopaulo.pinheiro@fqa.ub.edu}

\author[e]{Salvador Urrea}
\affiliation[e]{Instituto de F\'isica Corpuscular (IFIC), CSIC-UV,
  Edificio Institutos de Investigaci\'on, Calle Catedr\'atico Jos\'e
  Beltr\'an 2, E-46980 Paterna, Spain}

\abstract{We present a detailed analysis of the spectral data of
  Borexino Phase II, with the aim of exploiting its full potential to
  constrain scenarios beyond the Standard Model.  In particular, we
  quantify the constraints imposed on neutrino magnetic moments,
  neutrino non-standard interactions, and several simplified models
  with light scalar, pseudoscalar or vector mediators.  Our analysis
  shows perfect agreement with those performed by the collaboration on
  neutrino magnetic moments and neutrino non-standard interactions in
  the same restricted cases and expands beyond those, stressing the
  interplay between flavour oscillations and flavour non-diagonal
  interaction effects for the correct evaluation of the event rates.
  For simplified models with light mediators we show the power of the
  spectral data to obtain robust limits beyond those previously
  estimated in the literature.}

\preprint{IFT-UAM/CSIC-22-14, IFIC/22-15, FTUV-22-0404.7998,
  YITP-SB-2022-05}

\keywords{neutrinos, non-standard interactions, neutrino magnetic
  moment, light-mediators}

\begin{document}

\maketitle

\section{Introduction}

The scattering among electrons and neutrinos played an important
historical role in the establishment of the Standard Model (SM)
allowing for precise tests of the V-A structure of the weak
interactions.  Being a purely weak, purely leptonic, two-body
reaction, makes both the theoretical cross-section calculation and the
experimental signatures particularly clean.  The scattering of
$\nu_{\mu,\tau}$ on electrons is a purely weak neutral-current (NC)
process, while $\nu_e\, e^-$ scattering proceeds via both weak NC and
weak charged-current (CC) interactions.  The latter is of particular
interest because it is one of the few reactions for which the SM
predicts a large destructive interference between these two channels.
For the same reasons it is a sensitive probe of physics beyond the
Standard Model (BSM).  Paradigmatic examples include neutrino
electromagnetic properties, non-standard neutrino-matter interactions
(NSI), and new interactions mediated by light bosons.  Most precise
results on $\nu_\mu\, e^- \to \nu_\mu\, e^-$ were obtained by the
CHARM-II experiment~\cite{CHARM-II:1993phx}, while $\nu_e\, e^- \to
\nu_e\, e^-$ was first observed with an accelerator beam in the
experiment performed at LAMPF~\cite{Allen:1992qe}, followed by the
measurement at the LSND experiment~\cite{LSND:2001akn}.  Since the
pioneering work of Reines and collaborators~\cite{Reines:1976pv} the
scattering process $\bar\nu_e\, e^-\to \bar\nu_e\, e^-$ has been
studied at a greater level of precision using neutrinos produced in
nuclear reactors~\cite{Vidyakin:1992nf, Derbin:1993wy, MUNU:2005xnz},
most recently in the TEXONO~\cite{TEXONO:2006xds} and GEMMA
experiments~\cite{Beda:2012zz, Beda:2013mta}.  Neutrino electron
elastic scattering (ES) can also be searched for at experiments
designed to detect Coherent Elastic neutrino-Nucleus Scattering
(CE$\nu$NS): while the signal in the SM would be too low to yield a
significant number of events, the presence of BSM physics may lead to
much larger contributions.  In fact, CE$\nu$NS searches using reactor
neutrinos at the CONUS~\cite{CONUS:2020skt, Rink:2022rsx},
CONNIE~\cite{CONNIE:2019xid}, and Dresden-II reactor
experiments~\cite{Colaresi:2022obx, Coloma:2022avw} lead to very
strong constraints on some of the BSM scenarios outlined above.
Similar searches can be performed using pion decay-at-rest (DAR)
sources: for example, the data of the COHERENT experiment on
CsI~\cite{COHERENT:2017ipa} should be sensitive to an ES signal as
well~\cite{Coloma:2022avw} since it would pass their signal selection
cuts.

Neutrino electron elastic scattering is also key to the study of solar
neutrinos.  Solar neutrinos were first detected in real time through
ES scattering in water in Kamiokande~\cite{Kamiokande-II:1989hkh}.  In
fact, the observed rate of ES versus CC interactions allowed for the
first model-independent evidence of flavour transitions in solar
neutrinos in Super-Kamiokande and SNO~\cite{Super-Kamiokande:1998qwk,
  SNO:2001kpb}.  For the last 15 years the
Borexino~\cite{Borexino:2000uvj} experiment has been detecting ES of
solar neutrinos in liquid scintillator performing precision
spectroscopy of their energy spectrum at Earth~\cite{Borexino:2008dzn,
  Borexino:2017rsf, BOREXINO:2020aww}.  The main purpose of the
experiment was the independent determination of the fluxes produced by
the different thermonuclear reactions in the Sun.  Nevertheless, the
precision of their high-statistics spectral data also allows for
precision tests of ES and, as such, it has been used by the
collaboration itself~\cite{Borexino:2017fbd, Borexino:2019mhy}.
Borexino results have also been analyzed/adapted by phenomenological
groups to impose bounds on BSM scenarios (see, \textit{e.g.},
Refs.~\cite{Bilmis:2015lja, Harnik:2012ni, Agarwalla:2012wf,
  Khan:2019jvr, deGouvea:2009fp, Amaral:2020tga, Montanino:2008hu,
  Brdar:2017kbt, Khan:2017djo, Khan:2017oxw} for an incomplete list),
in general using either total event rates or early spectral data.

In this work we present a detailed analysis of the spectral data of
Borexino Phase II with the aim of exploiting its full potential to
constrain new physics which can significantly affect ES (such as
neutrino magnetic moments, NSI, and simplified models with light
scalar, pseudoscalar, and vector mediators).  In general BSM
scenarios, neutrino neutral-current interactions may be flavour
changing.  This brings the issue of the interplay between neutrino
oscillations and interactions which do not conserve the neutrino
flavour, which has not always been properly treated in previous
literature on this topic.  Section~\ref{sec:frameworks} presents the
formalism to correctly evaluate the event rates, accounting for
flavour oscillations in presence of flavour non-diagonal interactions
(details are also presented in Appendix~\ref{sec:appendix}).  This
section also introduces the interaction cross sections for the BSM
models considered.  Section~\ref{sec:analisis} contains the
description of our calculation of the expected spectra at Borexino and
the dataset used, as well as our treatment of backgrounds,
uncertainties and further details on the statistical analysis.  It is
complemented with Appendix~\ref{sec:borexdetails}.

The main results of our work are presented in Sec.~\ref{sec:results}.
As mentioned above, the Borexino collaboration has already performed
two analyses on BSM searches using a spectral fit of the Phase-II data
set: a first analysis which is used to set constraints on the neutrino
magnetic moment~\cite{Borexino:2017fbd}, and a later search for NSI
with electrons~\cite{Borexino:2019mhy}.  As we will show in
Sec.~\ref{sec:results}, our results are in excellent agreement with
those obtained in Ref.~\cite{Borexino:2017fbd} for the magnetic moment
scenario: this serves as validation for our $\chi^2$ construction and,
in particular, for the implementation of systematic uncertainties (a
key point in this analysis, given the large number of events collected
by Borexino).  Regarding NSI, our analysis expands over the results
obtained in Ref.~\cite{Borexino:2017fbd} in two main ways:
\textit{(i)} we include all NSI operators in neutrino flavour space at
once in the fit, thus allowing for potential correlation effects and
cancellations in the cross section; and \textit{(ii)} we consider four
different cases regarding the structure of the operators (vector,
axial vector, left-handed only, and right-handed only).  Finally, our
work includes the derivation of new constraints for simplified models
with new light vector, scalar, and pseudoscalar mediators, which to
the best of our knowledge have not been reported by the collaboration
yet.  Finally, we summarize our conclusions in Sec.~\ref{sec:conclu}.

\section{Theoretical Frameworks}
\label{sec:frameworks}

In the SM, the differential cross section for neutrino-electron ES
($\nu_\beta\, e^- \to \nu_\beta\, e^-$) can be expressed
as~\cite{Bahcall:1995mm}
\begin{multline}
  \label{eq:sm-elec}
  \dfrac{\dd\sigma^\text{SM}_\beta}{\dd T_e}(E_\nu, T_e)
  = \dfrac{2 G_F^2 m_e}{\pi}
  \bigg\lbrace
  c_{L\beta}^2 \Big[ 1 + \dfrac{\alpha}{\pi} f_-(y) \Big]
  + c_{R\beta}^2\, (1-y)^2 \Big[ 1 + \dfrac{\alpha}{\pi}f_+(y) \Big]
  \\
  - 2\, c_{L\beta}\, c_{R\beta}\, \dfrac{m_e y}{2E_\nu}
  \Big[ 1 + \dfrac{\alpha}{\pi} f_\pm(y) \Big]
  \bigg\rbrace \,,
\end{multline}
where $T_e$ stands for electron (kinetic) recoil energy, $E_\nu$ is
the neutrino energy, and $y \equiv T_e / E_\nu$, and $f_+$, $f_-$,
$f_\pm$ are loop functions given in Ref.~\cite{Bahcall:1995mm}, while
$\alpha$ stands for the fine-structure constant, $G_F$ is the Fermi
constant, and $m_e$ is the electron mass.  The effective couplings
$c_{L\beta}$ and $c_{R\beta}$ contain the contributions from the SM
NC, which are equal for all flavours, plus the CC contribution which
only affects $\nu_e e^-$ scattering:
\begin{equation}
  \label{eq:cm-coupl}
  \begin{aligned}
    c_{Le}
    &= \rho\,\Big[ \kappa_{e}(T_e)\sin^2\theta_w - \dfrac{1}{2} \Big] + 1 \,,
    &\quad
    c_{Re}
    &= \rho\, \kappa_{e}(T_e)\sin^2\theta_w \,,
    \\
    c_{L\tau} = c_{L\mu}
    &= \rho\, \Big[ \kappa_\mu(T_e)\sin^2\theta_w - \dfrac{1}{2} \Big] \,,
    &\quad
    c_{R\tau} = c_{R\mu}
    &= \rho\, \kappa_\mu(T_e)\sin^2\theta_w \,,
  \end{aligned}
\end{equation}
with $\rho$ and $\kappa_\beta(T_e)$ departing from $1$ due to
radiative corrections of the gauge boson self-energies and
vertices~\cite{Bahcall:1995mm}.  The minimum neutrino energy required
to produce an electron with a given recoil energy $T_e$ is $E_{\nu,
  \text{min}} = \big( T_e + \sqrt{T_e^2 + 2 \, m_e \, T_e}\, \big)
\big/ 2$, while the maximum kinetic energy for an electron produced by
a neutrino with given energy $E_\nu$ is $T_{e,\text{max}} = E_\nu^2
\big/ (E_\nu + m_e/2)$.  The corresponding cross section for
$\bar\nu_\beta \, e^-$ ES can be obtained from~\eqref{eq:sm-elec} with
the exchange $c_{L\beta}\leftrightarrow c_{R\beta}$.

We assume that scattering off electrons is incoherent, so for the
calculation of the number of events it is enough to multiply the total
cross section $\sigma^\text{SM}_\beta(E_\nu)$ --~obtained by
integrating Eq.~\eqref{eq:sm-elec} over $T_e$~-- by the overall number
of electrons in the fiducial volume.  In principle, a correction
factor taking into account that the target electrons are actually
bound to atoms have to be considered for recoil energies comparable to
atomic binding energies.  However, for the solar neutrino energies
considered here this effect can be safely neglected.

Solar neutrinos produced in the core of the Sun are purely $\nu_e$.
Propagation through vacuum and matter induces a non-trivial evolution
of their flavour, so that neutrinos reaching the detector are in
general not definite flavour eigenstates but rather a coherent
superposition of them, which can be described in terms of the elements
$S_{\beta e}$ of an unitary matrix $S$ (see
Appendix~\ref{sec:appendix}).
As discussed above, in the SM the ES cross section includes
contributions from both CC interactions (which are diagonal in the
flavour basis) and NC interactions (which are diagonal in \emph{any}
basis), so that the flavour basis is the most adequate to describe
them.  The common approach consists in projecting the neutrino state
at the detector over the flavour eigenstates, thus introducing the
$\nu_e\to \nu_\beta$ transition probabilities $P_{e\beta} \equiv
|S_{\beta e}|^2$, and then convoluting them with the cross sections
for pure flavour states given in Eq.~\eqref{eq:sm-elec}.  So the
number of interactions of solar neutrinos in Borexino will be
proportional to the \emph{probability-weighted} electron scattering
cross section:
\begin{equation}
  \label{eq:ES-prob}
  N_\text{ev} \propto
  \sum_\beta P_{e\beta}\, \sigma^\text{SM}_\beta \,.
\end{equation}
This well-known expression relies on the special role played in the SM
by the flavour basis, which as we have seen diagonalizes both CC and
NC interactions.  However, through this work we will consider several
BSM phenomenological scenarios leading to modifications of the
interaction cross section in Eq.~\eqref{eq:sm-elec}.  When the BSM is
not lepton-flavour conserving, the CC-defined flavour basis may no
longer correspond to the NC eigenstates, and extra care must be taken
with the interplay of oscillations and scattering.  To address this
issue, it is convenient to rewrite Eq.~\eqref{eq:ES-prob} in a
basis-independent form:
\begin{equation}
  \label{eq:ES-dens}
  N_\text{ev} \propto
  \Tr\Big[ \rho^{(e)}\, \sigma^\text{SM} \Big] \,,
\end{equation}
where $\rho^{(e)} \equiv S\, \Pi^{(e)} S^\dagger$ is the density
matrix of the neutrino state at the detector, with $\Pi^{(e)}$ being
the projector on the electron neutrino flavour at the source.  Here
$\sigma^\text{SM}$ is a \emph{generalized} cross section,
\textit{i.e.}, a matrix in flavour space which contains enough
information to describe the ES interaction of \emph{any} neutrino
state without the need to explicitly project it onto the interaction
eigenstates.  Such projection is now implicitly encoded into
$\sigma^\text{SM}$, which in the flavour basis takes the diagonal form
$\sigma^\text{SM}_{\gamma\beta} = \delta_{\gamma\beta}\,
\sigma^\text{SM}_\beta$.  In the same basis we have
$\rho_{\beta\gamma}^{(e)} = S_{\beta e} S_{\gamma e}^*$, so that
$\rho_{\beta\beta}^{(e)} = P_{e\beta}$ and Eq.~\eqref{eq:ES-prob} is
immediately recovered.  Let us stress that the indexes $\gamma\beta$
in $\sigma^\text{SM}_{\gamma\beta}$ do not make reference to
``initial'' and ``final'' states of the scattering process but they
are \emph{both} initial flavour indexes, just as the indexes of
$\rho_{\beta\gamma}^{(e)}$ both refer to the final state of the
neutrino as it reaches the detector.

\subsection{Neutrino Non-Standard Interactions with Electrons}

The so-called non-standard neutrino interaction (NSI) framework
consists on the addition of four-fermion effective operators to the SM
Lagrangian at low energies.  For example, the effective Lagrangian
\begin{equation}
  \label{eq:nsi-nc}
  \mathcal{L}_\text{NSI,NC} = -2\sqrt{2} G_F
  \sum_{f,P,\alpha,\beta} \Eps_{\alpha\beta}^{f,P}
  (\bar\nu_\alpha\gamma^\mu P_L\nu_\beta)
  (\bar f\gamma_\mu P f) \,,
\end{equation}
would lead to new NC interactions with the rest of the SM fermions.
Here $P$ can be either a left-handed or a right-handed projection
operator ($P_L$ or $P_R$, respectively) and the matrices
$\Eps_{\alpha\beta}^{f,P}$ are hermitian: $\big(
\Eps_{\beta\alpha}^{f,P} \big)^* = \Eps_{\alpha\beta}^{f,P}$.  The
index $f$ refers to SM fermions and in what follows we will be
focusing on $f=e$.  Such new interactions may induce lepton
flavour-changing processes (if $\alpha \neq \beta$), or may lead to a
modified interaction rate with respect to the SM result (if $\alpha =
\beta$).  For later convenience we also define the vector and axial SM
couplings:
\begin{equation}
  \label{eq:cva-coupl}
  c_{V\beta} \equiv c_{L\beta} + c_{R\beta} \,,
  \qquad
  c_{A\beta} \equiv c_{L\beta} - c_{R\beta} \,,
\end{equation}
as well as the corresponding NSI coefficients:
\begin{equation}
  \Eps_{\alpha\beta}^{f,V} \equiv
  \Eps_{\alpha\beta}^{f,L} + \Eps_{\alpha\beta}^{f,R} \,,
  \quad
  \Eps_{\alpha\beta}^{f,A} \equiv
  \Eps_{\alpha\beta}^{f,L} - \Eps_{\alpha\beta}^{f,R} \,.
\end{equation}

The presence of flavour-changing effects implies that the SM flavour
basis no longer coincides with the interaction eigenstates.  In such
case the number of events in Borexino will be given by
Eq.~\eqref{eq:ES-dens} in terms of a generalized cross section
$\sigma^\text{NSI}(E_\nu)$ defined as the integral over $T_e$ of the
following matrix expression:\footnote{Eq.~\eqref{eq:nsi-elec} is
already summed over the outgoing neutrino state, which is assumed to
be undetected. The expression for a specific final state $\nu_f$ can
be obtained by inserting the corresponding projector $\Pi^{(f)}$
between each pair of $C_{L,R}$ matrices, \textit{i.e.}, by replacing
$C_L^2 \to C_L \Pi^{(f)} C_L$, $C_R^2 \to C_R \Pi^{(f)} C_R$ and
$\big\{ C_L, C_R \big\} \to C_L \Pi^{(f)} C_R + C_R \Pi^{(f)}
C_L$. Since $\sum_f \Pi^{(f)} = I$, Eq.~\eqref{eq:nsi-elec} is
recovered upon summation over $f$.}
\begin{multline}
  \label{eq:nsi-elec}
  \dfrac{\dd\sigma^\text{NSI}}{\dd T_e}(E_\nu, T_e)
  = \dfrac{2 G_F^2 m_e}{\pi}
  \bigg\lbrace
  C_L^2 \Big[ 1 + \dfrac{\alpha}{\pi} f_-(y) \Big]
  + C_R^2\, (1-y)^2 \Big[ 1 + \dfrac{\alpha}{\pi}f_+(y) \Big]
  \\
  - \big\{ C_L, C_R \big\}\,
  \dfrac{m_e y}{2E_\nu} \Big[ 1 + \dfrac{\alpha}{\pi} f_\pm(y) \Big]
  \bigg\rbrace \,,
\end{multline}
which is formally identical to Eq.~\eqref{eq:sm-elec} except that the
$c_{L\beta}$ and $c_{R\beta}$ real coefficients have been superseded by
the $3\times 3$ hermitian matrices $C_L$ and $C_R$:
\begin{equation}
\label{eq:CL-CR}
  C_{\alpha\beta}^L
  \equiv c_{L\beta}\, \delta_{\alpha\beta} + \Eps_{\alpha\beta}^{e,L}
  \quad\text{and}\quad
  C_{\alpha\beta}^R
  \equiv c_{R\beta}\, \delta_{\alpha\beta} + \Eps_{\alpha\beta}^{e,R} \,.
\end{equation}
It is immediate to see that, if the NSI terms
$\Eps_{\alpha\beta}^{e,L}$ and $\Eps_{\alpha\beta}^{e,R}$ are set to
zero, the matrix $\dd\sigma^\text{NSI} \big/ \dd T_e$ becomes diagonal
and the SM expressions are recovered.  The corresponding cross section
for $\bar\nu_\beta \, e^-$ ES can be obtained from
Eq.~\eqref{eq:nsi-elec} with the exchange $C_L \leftrightarrow C_R^*$.

In addition to scattering effects, the presence of NSI also affects
neutrino propagation from production to detection point through its
contribution to the matter potential both in the Sun and in the Earth.
The formalism follows the approach described in Sec.~2.3 of
Ref.~\cite{Esteban:2018ppq}.  The details of the calculation of the
density matrix $\rho^{(e)}$ are provided in
Appendix~\ref{sec:appendix}.

\subsection{Neutrino Magnetic-Moments}

In the presence of a neutrino magnetic moment ($\mu_\nu$) the
scattering cross sections on electrons get additional contributions
which do not interfere with the SM ones.  Neutrino magnetic moments
arise in a variety of BSM models and, in particular, they do not need
to be flavour-universal.  Therefore, in what follows we will allow
different magnetic moments for the different neutrino flavours --
albeit still imposing that they are flavour-diagonal.  Under this
assumption the ES differential cross section for either neutrinos or
antineutrinos, up to order $\mathcal{O}(y^2)$, takes the
form~\cite{Vogel:1989iv}
\begin{equation}
  \label{eq:mag-elec}
  \frac{\dd\sigma^{\mu_\nu}_\beta}{\dd T_e}
  = \frac{\dd\sigma^\text{SM}_\beta}{\dd T_e}
  + \left(\frac{\mu_{\nu_\beta}}{\mu_B}\right)^2
  \frac{\alpha^2\,\pi}{m_e^2}
  \left[\frac{1}{T_e} - \frac{1}{E_\nu} \right] .
\end{equation}
Since no flavour-changing contributions are present in this case, the
expression for the number of interactions in Borexino will be just
proportional to the probability weighted total cross section, in
analogy with Eq.~\eqref{eq:ES-prob}.

\subsection{Models with light scalar, pseudoscalar, and vector mediators}

We also consider simplified models describing the neutrino interaction
with fermions of the first generation through a neutral scalar $\phi$,
pseudoscalar $\varphi$, or a neutral vector $Z'$.  Concerning the
scalar, the Lagrangian we consider is~\cite{Cerdeno:2016sfi}
\begin{equation}
  \label{eq:lagscal}
  \mathcal{L}_\phi = g_\phi \, \phi
  \bigg( q_\phi^e\, \bar{e} e + \sum_\alpha
  q_\phi^{\nu_\alpha}\, \bar\nu_{\alpha,R}\, \nu_{\alpha,L}
  + \text{h.c.} \bigg)
  - \frac{1}{2} \, M^2_\phi \, \phi^2 \,,
\end{equation}
where $q_\phi^j$ are the individual scalar charges and $j =
\{\nu_\alpha, e\}$.  The corresponding cross section for
neutrino-electron (or antineutrino-electron) scattering is
\begin{equation}
  \label{eq:csscale}
  \frac{\dd \sigma_{\beta}^\phi}{\dd T_e}
  = \frac{\dd\sigma^\text{SM}_\beta}{\dd T_e}
  + \frac{g_\phi^4 \, (q^{\nu_\beta}_\phi)^2 \, (q^e_\phi)^2 \, m_e^2 \, T_e}
  {4\pi\, E_\nu^2\, (2\, m_e\, T_e + M_\phi^2)^2}
  \left( 1 + \frac{T_e}{2\,m_e} \right) ,
\end{equation}
where we have included all orders in $T_e/E_\nu$ and $T_e/m_e$ (beyond
the leading-order terms given in Ref.~\cite{Cerdeno:2016sfi}), since
these are non-negligible for solar event rates at Borexino.

For the pseudoscalar scenario, we consider the Lagrangian
\begin{equation}
  \label{eq:lagpscal}
  \mathcal{L}_\varphi = i\, g_\varphi \, \varphi
  \bigg( q_\varphi^e\, \bar{e}\gamma^5 e + \sum_\alpha
  q_\varphi^{\nu_\alpha}\, \bar\nu_{\alpha,R} \gamma^5 \nu_{\alpha,L}
  + \text{h.c.} \bigg)
  - \frac{1}{2} \, M^2_\varphi \, \varphi^2 \,,
\end{equation}
where $q_\varphi^j$ are the individual pseudoscalar charges and $j =
\{\nu_\alpha, e\}$.  The corresponding cross section for
neutrino-electron (or antineutrino-electron) scattering is
\begin{equation}
  \label{eq:cspscale}
  \frac{\dd \sigma_{\beta}^{\varphi}}{\dd T_e}
  = \frac{\dd\sigma^\text{SM}_\beta}{\dd T_e}
  + \frac{g_\varphi^4 \, (q^{\nu_\beta}_\varphi)^2 \, (q^e_\varphi)^2 \, m_e \, T_e^2}
  {8 \pi \, E_\nu^2 \, (2\, m_e\, T_e + M_\varphi^2)^2} \,.
  \end{equation}

As for the vector mediator, our assumed Lagrangian is:
\begin{equation}
  \label{eq:lagvec}
  \mathcal{L}_V = g_{Z'}\, Z'_\mu \bigg( q_{Z'}^e\, \bar{e} \gamma^\mu e
  + \sum_\alpha q_{Z'}^{\nu_\alpha}\, \bar\nu_{\alpha,L}\gamma^\mu \nu_{\alpha,L}
 \bigg) + \frac{1}{2} M^2_{Z'} {Z'}^\mu Z'_\mu\,,
\end{equation}
with the individual vector charges $q_{Z'}^j$.  Unlike for the scalar
and magnetic-moment cases, a neutral vector interaction interferes
with the SM contribution.  The differential cross section for
neutrino-electron scattering reads~\cite{Lindner:2018kjo}
\begin{multline}
  \label{eq:csvece}
  \frac{\dd \sigma_{\beta}^{Z'}}{\dd T_e}
  = \frac{\dd\sigma^\text{SM}_\beta}{\dd T_e}
  + \frac{g^2_{Z'} \, m_e}{2 \, \pi} \Bigg\{
  \frac{g^2_{Z'}\, (q_{Z'}^{\nu_\beta})^2\, (q^e_{Z'})^2}{\big(2 m_e T_e + M_{Z'}^2\big)^2}
  \left[ 1 - \frac{m_e\, T_e}{2\,E_\nu^2}
    + \frac{T_e}{2\,E_\nu} \left( \frac{T_e}{E_\nu} - 2 \right) \right]
  \\
  + \frac{2\sqrt{2}\, G_F\, q_{Z'}^{\nu_\beta}\, q_{Z'}^e}{\big(2 m_e T_e + M_{Z'}^2\big)}
  \left[ c_{V\beta} \left( 1 - \frac{m_e\, T_e}{2\,E_\nu^2} \right)
    + c_{R\beta}\, \frac{T_e}{E_\nu} \left( \frac{T_e}{E_\nu} - 2 \right)
    \right]
  \Bigg\} \,,
\end{multline}
where $c_{V\beta}$ is the SM effective vector coupling introduced in
Eq.~\eqref{eq:cva-coupl}.  For the sake of simplicity, in our
calculations describing the effects of light mediators, we have
neglected the small SM radiative corrections to the BSM contributions.
The corresponding cross section for $\bar\nu_\beta \, e^-$ ES can be
obtained from~\eqref{eq:csvece} with the exchange
$c_{L\beta}\leftrightarrow c_{R\beta}$.

Similarly to the SM and the magnetic moment cases, no flavour-changing
contributions are considered for scalar, pseudoscalar, or vector
mediators, hence the number of interactions in Borexino will be just
proportional to the probability-weighted total cross section as in
Eq.~\eqref{eq:ES-prob}.

In Sec.~\ref{sec:results} we study the constraints on four specific
models with light mediators.  For scalar and pseudoscalar mediators we
will focus on a model in which the mediator couples universally to
fermions.  For vector mediators we will show the results for three
anomaly-free\footnote{We assume that three right-handed neutrinos are
added to the SM particle content which, in addition of generating
non-zero neutrino masses, cancel the anomalies in the $B-L$ case.}
models coupling to electrons, namely $B-L$, $L_e-L_\mu$ and
$L_e-L_\tau$.  For convenience Tab.~\ref{tab:charges} lists the
relevant charges.

\begin{table}\centering
  \catcode`?=\active \def?{\hphantom{+}}
  \begin{tabular}{l|cccc}
    Model & $q^e$ & $q^{\nu_e}$ & $q^{\nu_\mu}$ & $q^{\nu_\tau}$
    \\
    \hline
    Universal/leptonic scalar (or pseudoscalar) & $?1$ & $?1$ & $?1$ & $?1$
    \\
    $B-L$ vector & $-1$ & $-1$ & $-1$ & $-1$
    \\
    $L_e-L_\mu$ vector & $?1$ & $?1$ & $-1$ & $?0$
    \\
    $L_e-L_\tau$ vector & $?1$ & $?1$ & $?0$ & $-1$
  \end{tabular}
  \caption{Charges for the vector, scalar, and pseudoscalar mediators
    considered in this work.}
  \label{tab:charges}
\end{table}

\section{Analysis of Borexino Phase-II Spectrum}
\label{sec:analisis}

In order to constrain the BSM scenarios for neutrino interactions
described in Sec.~\ref{sec:frameworks} we have performed an analysis
of the Borexino Phase-II data collected between December 14th, 2011
and May 21st, 2016, (corresponding to a total exposure of
$1291.51~\text{days} \times 71.3~\text{tons}$) following closely the
details presented by the collaboration in
Ref.~\cite{Borexino:2017rsf}.  In particular we use the spectrum data
presented in Fig.~7 of such work which we analyze using the total
number of detected hits ($N_h$) on the detector photomultipliers
(including multiple hits on the same photomultiplier) as estimator of
the recoil energy of the electron scattered by the incoming neutrinos.
We statistically confront the spectral data with the expected event
rates in the different scenarios, which we compute as follows.

Borexino phase-II detects interactions of solar neutrinos produced in
the pp, pep, \Nuc{7}{Be}, and \Nuc{8}{B} reactions of the pp-chain, as
well as from all reactions in the CNO-cycle.  The produced total flux
is described by the differential spectrum:
\begin{equation}
  \label{eq:prod}
  \frac{\dd \phi_\nu^\text{prod}}{\dd E_\nu}
  = \sum_f \frac{\dd \phi_\nu^{f}}{\dd E_\nu}(E_\nu) ,
\end{equation}
where $E_\nu$ is the neutrino energy and $f$ refers to each of the
main components of the solar neutrino flux\footnote{Following the
analysis of Borexino we sum the fluxes from all reaction in the
CNO-cycle and treat them as a unique component.}  (pp, pep, CNO,
\Nuc{7}{Be}, and \Nuc{8}{B}).  In their journey to the Earth the
produced $\nu_e$ will change flavour and interact upon arrival via
elastic scattering with the $e^-$ of the detector.  Thus, the expected
number of solar neutrino events corresponding to $N_h$ hits is given
by the convolution of the oscillated solar neutrino spectrum with the
interaction cross section and the energy resolution function.
Quantitatively we compute the solar neutrino signal $S_i$ in the
$i$-th bin (\textit{i.e.}, with $N_h \in [N_{h,\text{min}}^i,
  N_{h,\text{max}}^i]$) as:
\begin{equation}
  \label{eq:binning}
  S_i = \int_{N_{h,\text{min}}^i}^{N_{h,\text{max}}^i}
  \int \frac{\dd S}{\dd T_e}(T_e)\,
  \frac{\dd\mathcal{R}}{\dd N_h}(T_e, N_h)\, \dd T_e\, \dd N_h \,.
\end{equation}
Here, $\dd S \big/ \dd T_e$ is the differential distribution of
neutrino-induced events as a function of recoil energy of the
scattered electrons ($T_e$)
\begin{equation}
  \label{eq:dSdTe}
  \frac{\dd S}{\dd T_e}(T_e)
  = \mathcal{N}_\text{tgt}\, \mathcal{T}_\text{run}\,
  \mathcal{E}_\text{cut}
  \int \frac{\dd \phi_\nu^\text{prod}}{\dd E_\nu}(E_\nu)
  \Tr\left[ \rho^{(e)}(E_\nu)\,
    \frac{\dd \sigma^{X}}{\dd T_e}(E_{\nu}, T_e) \right]
  \dd E_{\nu} \,,
\end{equation}
where $X \in \{ \text{SM, NSI, $Z'$, $\phi$, $\varphi$, $\mu_\nu$} \}$
for the different scenarios discussed in Sec.~\ref{sec:frameworks}.
Note that our notation explicitly reflects that $\dd\sigma \big/ \dd
T_e$ may not be diagonal in flavour space (as in the case of NSI, see
the discussion in Sec.~\ref{sec:frameworks}).  In our analysis we fix
the oscillation parameters to the best-fit values from
Ref.~\cite{Esteban:2020cvm, nufit}.  Here $\mathcal{N}_\text{tgt}$ is
the number of $e^-$ targets, that is, the total number of electrons
inside the fiducial volume of the detector (corresponding to 71.3~ton
of scintillator), while $\mathcal{T}_\text{run} = 1291.51~\text{days}$
is the data taking time, and $\mathcal{E}_\text{cut} = 98.5\%$ is the
overall efficiency.
In addition Eq.~\eqref{eq:binning} includes the energy resolution
function $\dd\mathcal{R} \big/ \dd T_e$ for the detector which gives
the probability that an event with electron recoil energy $T_e$ yields
an observed number of hits $N_h$.  We assume it follows a Gaussian
distribution
\begin{equation}
  \label{eq:dist-diff}
  \frac{\dd\mathcal{R}}{\dd N_h}(T_e, N_h)
  = \frac{1}{\sqrt{2\pi} \, \sigma_h(T_e)}
  \exp\bigg[ -\frac{1}{2} \bigg(
    \frac{N_h - \bar{N}_h(T_e)}{\sigma_h(T_e)} \bigg)^2 \bigg] \,,
\end{equation}
where $\bar{N}_h$ is the expected value of $N_h$ for a given true
recoil energy $T_e$.  We determine $\bar{N}_h(T_e)$ and
$\sigma_h(T_e)$ via the calibration procedure described in
Appendix~\ref{sec:calibration}.

The data contains background contributions from a number of sources.
The main backgrounds come from radioactive isotopes in the
scintillator \Nuc{14}{C}, \Nuc{11}{C}, \Nuc{10}{C}, \Nuc{210}{Po},
\Nuc{210}{Bi}, \Nuc{85}{Kr}, and \Nuc{6}{He}.  The collaboration
identifies two additional backgrounds due to pile-up of uncorrelated
events, and residual external backgrounds (see
Ref.~\cite{Borexino:2017rsf} for a complete description of these
backgrounds).

Among the considered backgrounds, the one coming from \Nuc{11}{C} is
particularly relevant for the analysis.  This isotope is produced in
the detector by muons through spallation on \Nuc{12}{C}.  In
Ref.~\cite{Borexino:2017rsf} the collaboration uses a Three-Fold
Coincidence (TFC) method to tag the \Nuc{11}{C} events, which are
correlated in space and time with a muon and a neutron.  Thus, they
divide the Phase II data set in two samples: one enriched (tagged) and
one depleted (subtracted) in \Nuc{11}{C} events.  The separation of
the solar neutrino signal (as well as most of the other backgrounds)
into these two samples is uncorrelated from the number of hits
$N_h$. Concretely, the tagged sample picks up $35.72\%$ of the solar
neutrino events, while the subtracted sample accounts for the
remaining $64.28\%$.

In our analysis we include both sets of data, denoted in what follows
by $s = \text{``tagged''}$ or ``subtracted''.  As for the background,
we have read the contribution $B_{s,i}^c$ for each component $c$, in
each bin $i$, and for each data set $s$ from the corresponding lines
in the two panels in Fig.~7 of Ref.~\cite{Borexino:2017rsf}.  They are
shown for the best-fit normalization of the different background
components as obtained by the collaboration, and we take them as our
nominal background predictions in the absence of systematics.
Altogether the nominal number of expected events $T_{s,i}^0$ in some
bin $i$ of data sample $s$ is the sum of the neutrino-induced signal
and the background contributions:
\begin{equation}
  T_{s,i}^0 = \sum_f S_{s,i}^f  + \sum_c B_{s,i}^c \,,
\end{equation}
where the index $f \in \{ \text{pp, \Nuc{7}{Be}, pep, CNO, \Nuc{8}{B}}
\}$ runs over the solar fluxes, while the index $c \in \{
\text{\Nuc{14}{C}, \Nuc{11}{C}, \Nuc{10}{C}, \Nuc{210}{Po},
  \Nuc{210}{Bi}, \Nuc{85}{Kr}, \Nuc{6}{He}, pile-up, ext} \}$ runs
over the background components.  In our calculation of $S_{s,i}^f$ we
use the high-metallicity (HZ) solar model for simplicity.\footnote{It
should be noted that this is the model currently favoured by the CNO
measurement at Borexino~\cite{BOREXINO:2020aww}, albeit with a modest
significance.}

\begin{figure}[t]\centering
  \includegraphics[width=0.95\textwidth]{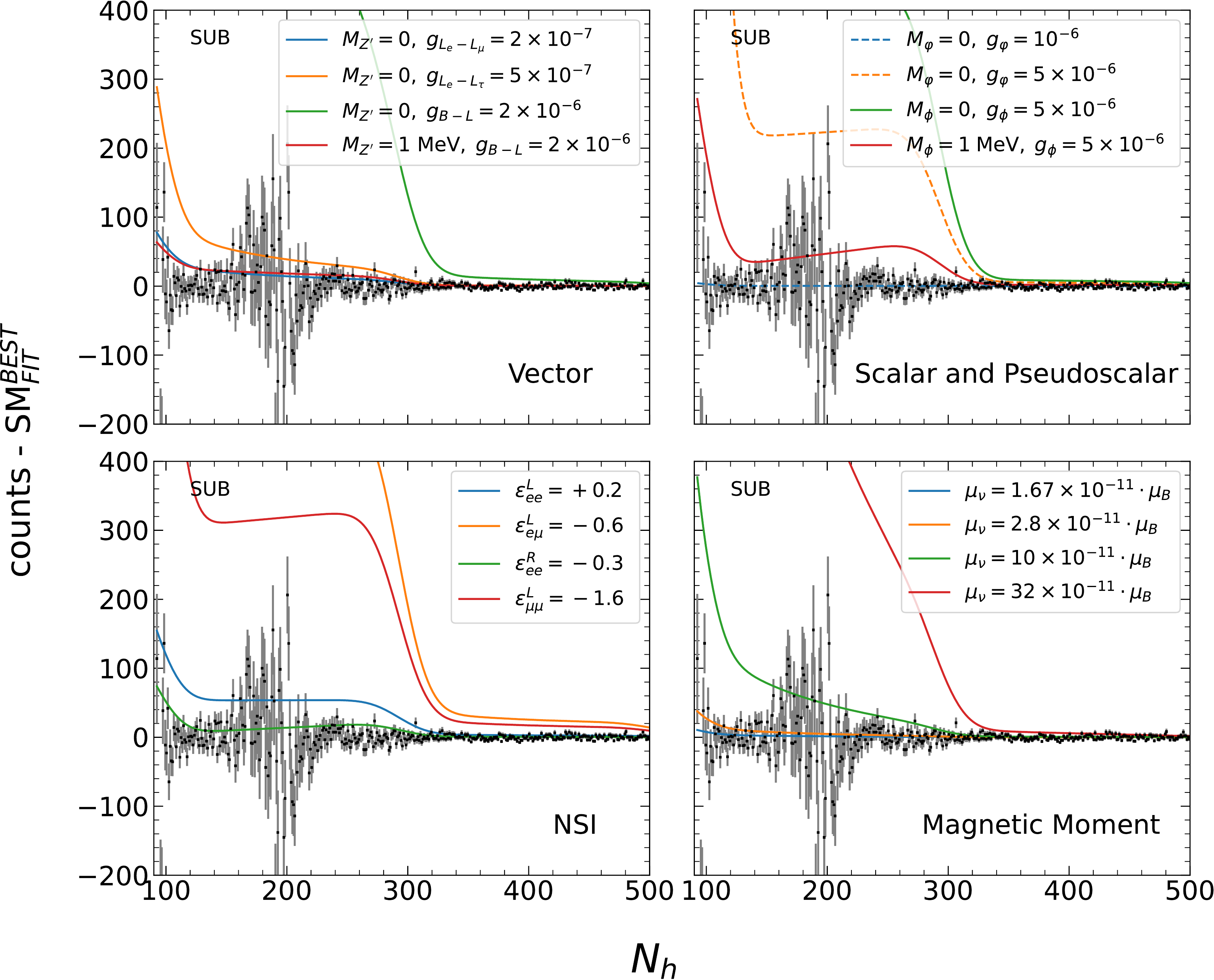}
  \caption{Difference between the number of events with respect to the
    SM expectation, as a function of $N_h$.  For reference, the black
    dots are obtained for the data in the ``subtracted'' sample, where
    the error bars indicate statistical errors only.  The coloured
    lines correspond to the predicted event rates (using the same
    exposure as that of the ``subtracted'' sample) for different BSM
    scenarios and parameter values, as indicated by the labels.}
  \label{fig:events}
\end{figure}

In Fig.~\ref{fig:events} we show the predicted event distributions for
some examples of the new physics models we consider.  Along with these
spectra, we also include the Borexino data points, after subtracting
the best-fit background model and SM signal.  For the sake of
concreteness we show the results for the ``subtracted'' event sample.
As seen in the figure, the BSM scenarios considered here yield larger
contribution in the low energy bins.  Therefore the derivation of
robust bounds requires a detailed treatment of the background
subtraction and careful accounting of all systematic uncertainties, in
particular those associated with the energy reconstruction.  This was
also the case for the analysis performed by the Borexino collaboration
in Ref.~\cite{Borexino:2017rsf} for the determination of the solar
fluxes, in particular the pp flux.  We follow closely such analysis in
our construction of the $\chi^2$ function as we outline next.  We
provide a detailed description of our treatment of backgrounds and
systematic uncertainties in Appendix~\ref{sec:borexdetails}.

Our statistical analysis is based on the construction of a $\chi^2$
function based on the described experimental data, neutrino signal
expectations and sources of backgrounds, as well as the effect of
systematic uncertainties.  We introduce the latter in terms of
\emph{pulls} $\xi_r$, which are then varied in order to minimize the
$\chi^2$.  The analysis includes a total of 27 pulls to account for
the uncertainties in the background normalizations, detector
performance, priors on the solar fluxes, and energy shifts (see
Appendix~\ref{sec:borexdetails} for details).  Assuming a linear
dependence of the event rates on the pulls, the minimization can be
performed analytically provided the $\chi^2$ can be assumed to be
Gaussian.  Borexino collaboration bins their data in each sample using
858 bins, from $N_h=92$ to $N_h=950$.  Such thin binning does not
provide enough events in each bin to guarantee Gaussian statistics.
So for convenience we rebin the data using coarser bins: the first 42
bins are rebinned in groups of 2, the next 210 bins in groups of 5,
the next 260 bins in groups of 10 and the last 346 bins in groups of
50.  This yields a total of 96 bins $k$ for each data sample, $s$,
each of them having enough events to guarantee Gaussian statistics.
Altogether the $\chi^2$ we employ reads:
\begin{equation}
  \label{eq:chi2}
  \chi^2 = \min_{\vec\xi} \Bigg[ \sum_{s,k}
    \frac{\big[ T_{s,k}(\vec\xi) - O_{s,k} \big]^2}{O_{s,k}}
    + \sum_r (\xi^{\text{unc}}_r)^2 + \sum_{r,r'} (\Sigma^{-1})_{rr'}\,
    \xi_r^\text{corr}\, \xi_{r'}^\text{corr} \Bigg] \,,
\end{equation}
where $O_{s,k}$ is the observed number of events in bin $k$ of sample $s$
and $T_{s, k}$ is the corresponding  predicted  number of events in the
presence of systematics
\begin{equation}
  \label{eq:tsys}
  T_{s,k}(\vec\xi) = T_{s,k}^0 + \sum_r D_{s,k}^r \xi_r \,.
\end{equation}
$D_{s, k}^{r}$ stands for the derivatives of the total event rates
with respect to each pull, which are given in
Appendix.~\ref{sec:borexdetails}.

The last two terms in Eq.~\eqref{eq:chi2} are Gaussian bias factors
which we include for those pulls for which some prior constraint
should be accounted for.  The first of those last terms correspond to
pulls associated to uncorrelated (``unc'') uncertainties while the
last one to correlated (``corr'') uncertainties.  Here, $\Sigma$
stands for the covariance matrix used for the latter and it is given
in Appendix~\ref{sec:borexdetails}.

\section{Results}
\label{sec:results}

This section summarizes our results from the spectral analysis of
Borexino data, as outlined in Sec.~\ref{sec:analisis}, for the
phenomenological scenarios defined in Sec.~\ref{sec:frameworks}.  We
will first present our analysis for the magnetic moment scenario
introduced in Sec.~\ref{sec:magnetic}.  As we will see, our results
are in excellent agreement with those obtained in
Ref.~\cite{Borexino:2017fbd} and therefore serve to validate our
$\chi^2$ implementation (and, in particular, the implementation of
systematic uncertainties) discussed in Sec.~\ref{sec:analisis}.  We
then proceed to the NSI scenario in Sec.~\ref{sec:NSI}, while the
cases of light vector, scalar, and pseudoscalar mediators are shown in
Secs.~\ref{sec:light-vec} and~\ref{sec:light-phi}.

At this point it is worth stressing the technical challenges
associated to the $\chi^2$ minimization procedure. The analysis, a
priori, depends on 27 pulls, 6 standard oscillation parameters, plus a
number of non-standard oscillation parameters (ranging from 2 to 6,
depending on the particular scenario considered). In order to make the
problem tractable from the computational point of view, we minimize
analytically over the 27 pulls as outlined in the previous section,
and we fix the oscillation parameters to the present best-fit values
from Ref.~\cite{Esteban:2020cvm, nufit} (we have checked that this has
a negligible impact on the results). The exploration of the BSM
parameter space is performed numerically, using the Diver
software~\cite{Martinez:2017lzg, diver}. We have also cross-checked
that the resulting two-dimensional confidence regions agree with the
result obtained using a simplex method algorithm.

Throughout this section, we will follow a frequentist approach in
order to determine the allowed confidence regions for the parameters
in each case.  In other words, for a given BSM scenario characterized
by a set of parameters $\vec\omega$, we define a $\Delta\chi^2$
function as:
\begin{equation}
  \Delta\chi^2(\vec\omega)
  = \chi^2(\vec\omega) - \chi^2_\text{min} \,,
\end{equation}
where $\chi^2_\text{min}$ refers to the global minimum of the $\chi^2$
function, found after minimization over all the model parameters
$\vec\omega$.  We stress that $\chi^2(\vec\omega)$ corresponds to
Eq.~\eqref{eq:chi2} where $T_{s,k}(\vec\xi)$ is in fact
$T_{s,k}(\vec\omega,\, \vec\xi)$ so the minimization over the pull
parameters is performed for each given value of the model
parameters.\footnote{Here it should be stressed that our allowed
regions for light mediators will be derived on the coupling, for
\emph{fixed} values of the mediator mass.  That is, the minimum value
of the $\chi^2$ is searched for each value of the mediator mass
independently, and the $\Delta\chi^ 2$ for that mass is defined with
respect to the minimum found.  Thus, the reported regions will be
derived assuming that the test statistic is distributed as a $\chi^2$
with one degree of freedom, instead of two.}  The allowed regions are
then determined by the set of model parameters for which the
$\Delta\chi^2$ lies below a certain level, as indicated in each case.

For reference let us mention that the analysis performed within the SM
results into $\chi^2_\text{min,SM} = 256.7$ for a total of 192 data
points and 27 pull parameters.

\subsection{Neutrino magnetic moment}
\label{sec:magnetic}

\begin{figure}[t]\centering
  \includegraphics[width=0.85\textwidth]{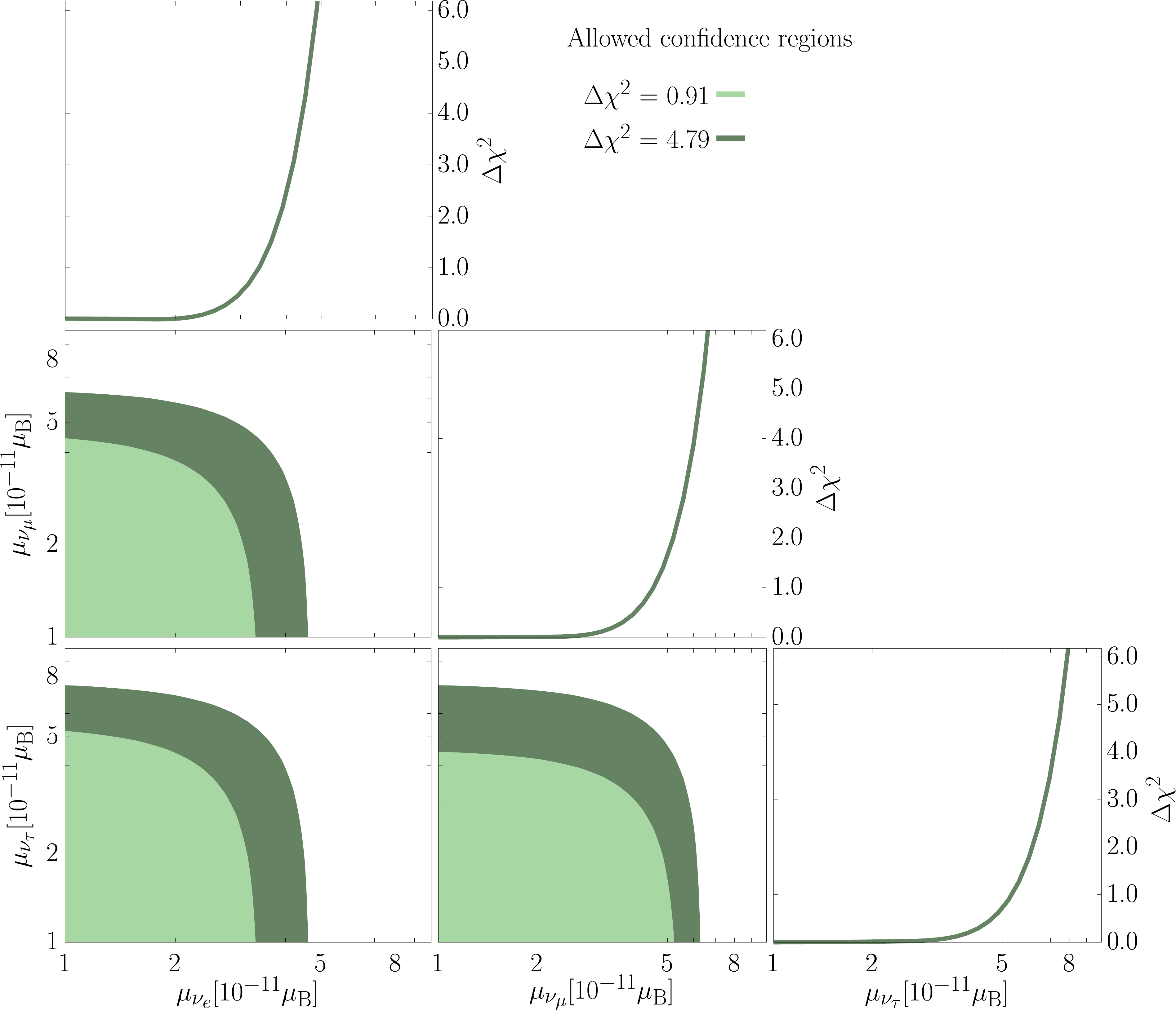}
  \caption{Results for the magnetic moment scenario, assuming three
    independent parameters (one for each of the SM neutrino flavours).
    Since the neutrino magnetic moment is positive definite, the cuts
    have been taken according to a one-sided $\chi^2$ distribution.
    In the central panels, we show the allowed confidence regions (at
    $1\sigma$ and $2\sigma$, for 2 d.o.f.\ using one-sided intervals)
    while in the right-most panels we show the profile of the
    $\Delta\chi^2$ for each of the parameters individually.  In each
    panel, the results are obtained after minimization over the
    undisplayed parameters.}
  \label{fig:magnetic}
\end{figure}

For the analysis with neutrino magnetic moment we find
$\chi^2_\mathrm{min,SM+\mu_\nu} = 256.5$.  In other words, the data do
not show a statistically significant preference for a non-vanishing
magnetic moment, which allows to set constraints on the parameter
space for this scenario.  The corresponding results are summarized in
Fig.~\ref{fig:magnetic}, assuming three independent parameters for the
$\nu_e$, $\nu_\mu$ and $\nu_\tau$ magnetic moments.  In each panel,
the $\Delta\chi^2$ is minimized over the undisplayed parameters.
From the one-dimensional projections we can read the upper bounds at
90\% CL (for 1 d.o.f., one-sided intervals, \textit{i.e.},
{$\Delta\chi^2 = 1.64$})
\begin{equation}
  \label{eq:mubounds}
  \mu_{\nu_e} < 3.7 \times 10^{-11} \mu_B \,,
  \qquad
  \mu_{\nu_\mu} < 5.0 \times 10^{-11} \mu_B \,,
  \qquad
  \mu_{\nu_\tau} < 5.9 \times 10^{-11} \mu_B \,.
\end{equation}
Our results are fully compatible with those in
Ref.~\cite{Borexino:2017fbd}, considering the different choice of
oscillation parameters assumed.  If we assume that all flavours carry
the same value of the neutrino magnetic moment then the dependence on
the oscillation parameters cancels out and we obtain the following
bound (at 90 \% CL for 1 d.o.f., one-sided)
\begin{equation}
  \label{eq:mubounds2}
  \mu_\nu < 2.8 \times 10^{-11} \mu_B \,.
\end{equation}
This coincides precisely with the result obtained by the collaboration
in Ref.~\cite{Borexino:2017fbd}, and serves as validation for the
$\chi^2$ implementation and the choice of systematic uncertainties
adopted in our analysis.

We notice that the Borexino bounds are stronger than those obtained by
other experiments under the same flavour assumptions and CL.  This is
the case for the (over a decade old) experimental results of
GEMMA~\cite{Beda:2010hk, Beda:2012zz} ($\mu_\nu < 2.9 \times 10^{-11}
\mu_B$) and TEXONO~\cite{TEXONO:2006xds} ($\mu_\nu < 7.4 \times
10^{-11} \mu_B$), as well as the bounds derived from recent results
such as CONUS~\cite{CONUS:2022qbb} ($\mu_\nu < 7.5 \times 10^{-11}
\mu_B$) and the combined analysis of Dresden-II and
COHERENT~\cite{Coloma:2022avw} ($\mu_\nu < 1.8 \times 10^{-10}
\mu_B$).

\subsection{Non-Standard neutrino Interactions with electrons}
\label{sec:NSI}

Let us now discuss the case of NSI with electrons.  In this scenario
Borexino is sensitive to a total of 12 operators: 6 operators
involving left-handed electrons (with coefficients
$\Eps_{\alpha\beta}^L$) and 6 operators involving right-handed
electrons (with coefficients $\Eps_{\alpha\beta}^R$), see
Eq.~\eqref{eq:nsi-nc}.  In all generality, \emph{a priori} all
operators should be considered at once in the analysis since all of
them enter at the same order in the effective theory.  However,
including such a large number of parameters makes the problem
numerically challenging, and the extraction of meaningful conclusions
is also jeopardized by the multiple interference effects between the
different coefficients.  Thus, here we focus on four representative
cases with a reduced set of operators, depending on the type of
interaction that could lead to the terms in Eq.~\eqref{eq:nsi-nc}: (1)
the axial vector case, where both left- and right-handed NSI operators
are generated with equal strength but with opposite signs; (2) the
purely vector case, where both types of operators are generated with
equal strength and equal sign; and (3) the purely left- or purely
right-handed cases, where only operators involving electrons of a
given chirality are obtained.  For each of these four benchmarks we
include all operators in neutrino flavour space (all flavour-diagonal
and off-diagonal operators) simultaneously in the analysis, allowing
for interference and correlation effects among them.

\begin{figure}[t]\centering
  \includegraphics[width=0.98\textwidth]{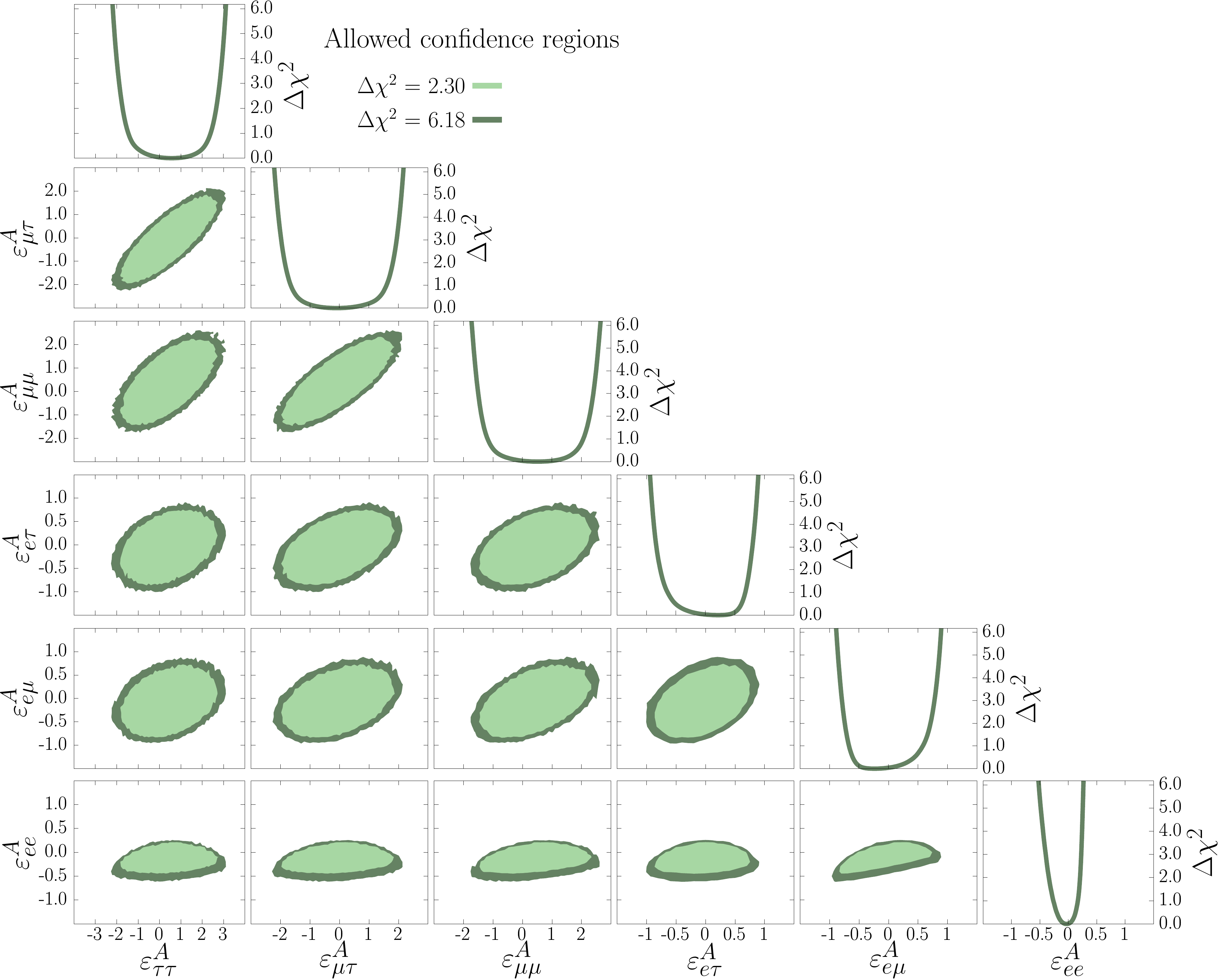}
  \caption{Results for axial NSI with electrons ($\Eps_{\alpha\beta}^L
    = -\Eps_{\alpha\beta}^R$).  The central panels show the
    two-dimensional allowed confidence regions at $1\sigma$ and
    $2\sigma$ (for 2 d.o.f., using two-sided intervals $\Delta\chi^2 =
    2.30$, $6.18$), while the remaining panels show the profile of the
    $\Delta\chi^2$ for each of the parameters individually.  In each
    panel, the results are obtained after minimization over the
    parameters not shown.}
  \label{fig:NSIcorner-a}
\end{figure}

We start discussing the axial NSI case.  For axial NSI, the effect
cancels out in the matter potential, rendering neutrino oscillations
insensitive to the new interactions.  However, Borexino can still
constrain these operators since they affect the interaction cross
section in the detector.  First, we notice that even allowing all 6
axial NSI coefficients to vary freely we get
$\chi^2_\mathrm{min,SM+NSI^A} = 255.7$, which is only one unit lower
that the SM.  So we proceed to derive the bounds on the full parameter
space.

The results are presented in Fig.~\ref{fig:NSIcorner-a} where we show
one-dimensional and two-dimensional projections of
$\Delta\chi^2_\mathrm{SM+NSI^A}$ after minimization over the
parameters not shown in each panel.  We notice that since the
cross-section of flavour-conserving interactions receive contributions
from both SM and BSM terms, which could lead to cancellations among
them, the corresponding regions are generically larger than those
involving only flavour-off-diagonal NSI, for which no SM term is
present.  From the figure we see that when all the parameters are
included in the fit, Borexino is able to constrain NSI roughly only at
the level of $|\Eps^A_{\alpha\alpha}| < \mathcal{O}(2)$ for
flavour-diagonal interactions, and $|\Eps^A_{\alpha\neq\beta}| <
\mathcal{O}(1)$ for the flavour-off-diagonal ones (see also
Tab.~\ref{tab:VA}).  The only exception to this is the parameter
$\Eps_{ee}^A$, for which tighter constraints are obtained, roughly at
the level $|\Eps_{ee}^A|< \mathcal{O}(0.5)$ as it can be seen from the
lowest row in the figure.  The main reason behind this is that the
$\Eps_{ee}^A$ parameter enters the effective cross section in
Eq.~\eqref{eq:nsi-nc} multiplied by $P_{ee}\sim 0.5$, while the
remaining NSI parameters are multiplied either by $P_{e\mu}, P_{e\tau}
\sim 0.2 - 0.25$ or by off-diagonal elements in the density matrix
(which are at most $\sim 0.2$).

Note also that the allowed regions for flavour-diagonal NSI are a bit
asymmetric with respect to zero.  In general, given the large
parameter space it is difficult to identify a single reason for the
specific shape of the allowed regions.  Still, the origin of the
asymmetry can be qualitatively understood if the generalized cross
section matrix in Eq.~\eqref{eq:nsi-elec} is written in terms of the
axial and vector coefficients, defined (in analogy to
Eq.~\eqref{eq:CL-CR}) as
\begin{equation}
  \begin{aligned}
    C^V_{\alpha\beta}
    &= C^L_{\alpha\beta} + C^R_{\alpha\beta} =
    c_{V\beta}\delta_{\alpha\beta} + \Eps^V_{\alpha\beta} \,,
    \\
    C^A_{\alpha\beta}
    &= C^L_{\alpha\beta} - C^R_{\alpha\beta} =
    c_{A\beta}\delta_{\alpha\beta} + \Eps^A_{\alpha\beta} \,.
  \end{aligned}
\end{equation}
In this case, the differential cross section for axial NSI, setting
$\alpha=\beta$, depends on the following combination of parameters:
\begin{equation}
  \label{eq:nsi-cv-ca}
  \frac{1}{2}\left(c_{V \alpha}^2 + c_{A \alpha}^2\right) + c_{A
    \alpha}\Eps_{\alpha\alpha}^A + \frac{(\Eps^A_{\alpha\alpha})^2}{2}
  + \mathcal{O}(y) \,.
\end{equation}
We notice that due to the interference between the SM and NSI
parameters, a second \emph{SM-like} solution appears in
Eq.~\eqref{eq:nsi-cv-ca} with $\Eps^A_{\alpha\alpha} \simeq -2\,
c_{A\alpha}$ (\textit{i.e.}, for which $C^A_{\alpha\alpha} =
-c_{A,\alpha}$).  It can actually be shown that for $\alpha = \mu$ and
$\alpha = \tau$ such \emph{SM-like} solution holds also after
including the full $y$ dependence of the cross section and so it
provides a fit of similar quality to the SM solution.  When only one
NSI parameter is considered at a time this leads to the two
disconnected allowed ranges seen for $\Eps^A_{\mu\mu}$ and
$\Eps^A_{\tau\tau}$ in Table~\ref{tab:VA}.  When all NSI parameters
are included both allowed regions merge in a unique range which is
asymmetric about zero.  In the case of $\Eps^A_{ee}$, on the contrary,
the $y$-dependent terms break such degeneracy and the second
\emph{SM-like} solution gets lifted.

From Eq.~\eqref{eq:nsi-cv-ca} we also see that for
$\Eps^A_{\alpha\alpha} \simeq -c_{A\alpha}$ it is possible to reduce
the cross section with respect to the SM case.  This leads to a
slightly better fit to the data (albeit mildly, at the level of
$\Delta\chi^2 = -1$ as mentioned above) and results into the preferred
solution to be at $\Eps_{ee}^A \lesssim 0$ (since $c_{Ae} \sim 0.5$).
Conversely, for $\Eps_{\mu\mu}^A$ and $\Eps_{\tau\tau}^A$ the fit
shows a mild preference for positive values since $c_{A\mu} =
c_{A\tau} \sim -0.5$.

Finally, it is worth stressing the impact of correlations in the fit,
which are large for this scenario.  For example, from the
two-dimensional panels in Fig.~\ref{fig:NSIcorner-a} it is evident
that there is a strong correlation between $\Eps^A_{\mu\tau}$,
$\Eps^A_{\mu\mu}$, and $\Eps^A_{\tau\tau}$, which has a significant
impact on the final limits obtained when all parameters are allowed to
float freely in the fit.  This is shown explicitly in
Tab.~\ref{tab:VA}, which compares the obtained bounds (at 90\% CL) for
the different NSI parameters in two different cases: turning on only
one parameter at a time (``1 Parameter''), or including all operators
simultaneously in the fit (``Marginalized'').  As can be seen from the
comparison between the two sets of results in this table, the impact
due to correlations and degeneracies between different operators is
significant and leads to considerably larger allowed regions if
multiple NSI operators are included at once.  For example, in the case
of $\Eps_{\mu\mu}^A$, the 90\% CL range is reduced by a factor of
$\sim 8$ if the remaining NSI operators are set to zero.

\begin{table}[t]\centering
  \renewcommand{\arraystretch}{1.2}
  \begin{tabular}{|c || c | c ||c | c ||}
    \hline
    & \multicolumn{4}{|c||}{Allowed regions at 90\% CL $(\Delta\chi^2 = 2.71)$}
    \\
    \cline{2-5}
    & \multicolumn{2}{|c||}{Vector}
    & \multicolumn{2}{|c||}{Axial Vector}
    \\
    \cline{2-5}
    & 1 Parameter & Marginalized
    & 1 Parameter & Marginalized
    \\
    \hline
    $\Eps_{ee}$
    & $[-0.09, +0.14]$ & $[-1.05, +0.17]$
    & $[-0.05, +0.10]$ & $[-0.38, +0.24]$
    \\
    $\Eps_{\mu\mu}$
    & $[-0.51, +0.35]$ & $[-2.38, +1.54]$
    &  $[-0.29, +0.19] \oplus [+0.68, +1.45]$ & $[-1.47, +2.37]$
    \\ $\Eps_{\tau\tau}$
    & $[-0.66, +0.52]$ & $[-2.85, +2.04]$
    & $[-0.40, +0.36] \oplus [+0.69, +1.44]$ & $[-1.82, +2.81]$
    \\
    $\Eps_{e\mu}$
    & $[-0.34, +0.61]$ & $[-0.83, +0.84]$
    & $[-0.30, +0.43]$ & $[-0.79, +0.76]$
    \\
    $\Eps_{e\tau}$
    & $[-0.48, +0.47]$ & $[-0.90, +0.85]$
    & $[-0.40, +0.38]$ & $[-0.81, +0.78]$
    \\
    $\Eps_{\mu\tau}$
    & $[-0.25, +0.36]$ & $[-2.07, +2.06]$
    & $[-1.10, -0.75] \oplus [-0.13, +0.22]$ & $[-1.95, +1.91]$
    \\
    \hline
  \end{tabular}
  \caption{90\% CL bounds (1 d.o.f., 2-sided) on the coefficients of
    NSI operators with electrons, for the vector ($\Eps^V$) and axial
    vector ($\Eps^A$) scenarios.  Results are provided separately for
    two cases: when only one NSI operator is included at a time (``1
    Parameter'') or when the remaining NSI coefficients are allowed to
    float freely in the fit (``Marginalized'').}
  \label{tab:VA}
\end{table}

Let us now turn to the analysis of vector NSI.  In this case NSI
affect the matter potential felt by neutrinos in propagation and a
relevant question is how sensitive the analysis is to this effect, and
whether the sensitivity to NSI is still dominated by the impact on the
neutrino detection cross section.  We have numerically checked that
the results of the analysis are rather insensitive to NSI effects in
propagation.  This is partly so because in the energy window
considered here the contribution from the higher-energy components
(mainly \Nuc{8}{B}, but also pep neutrinos) to the event rates is
subdominant, and moreover it is affected by relatively large
backgrounds uncertainties.  As a result we find that the impact of the
matter potential on the fit is very limited and the sensitivity comes
dominantly from NSI effects on the detection cross section.\footnote{A
similar finding was also reported in Ref.~\cite{Borexino:2019mhy},
although only on-diagonal operators were considered in that case.}  We
also find that, even allowing all 6 vector NSI coefficients to vary
freely in the analysis, $\chi^2_\mathrm{min,SM+NSI^V} -
\chi^2_\mathrm{min,SM}$ is only $-0.2$.

\begin{figure}[t]\centering
  \includegraphics[width=0.98\textwidth]{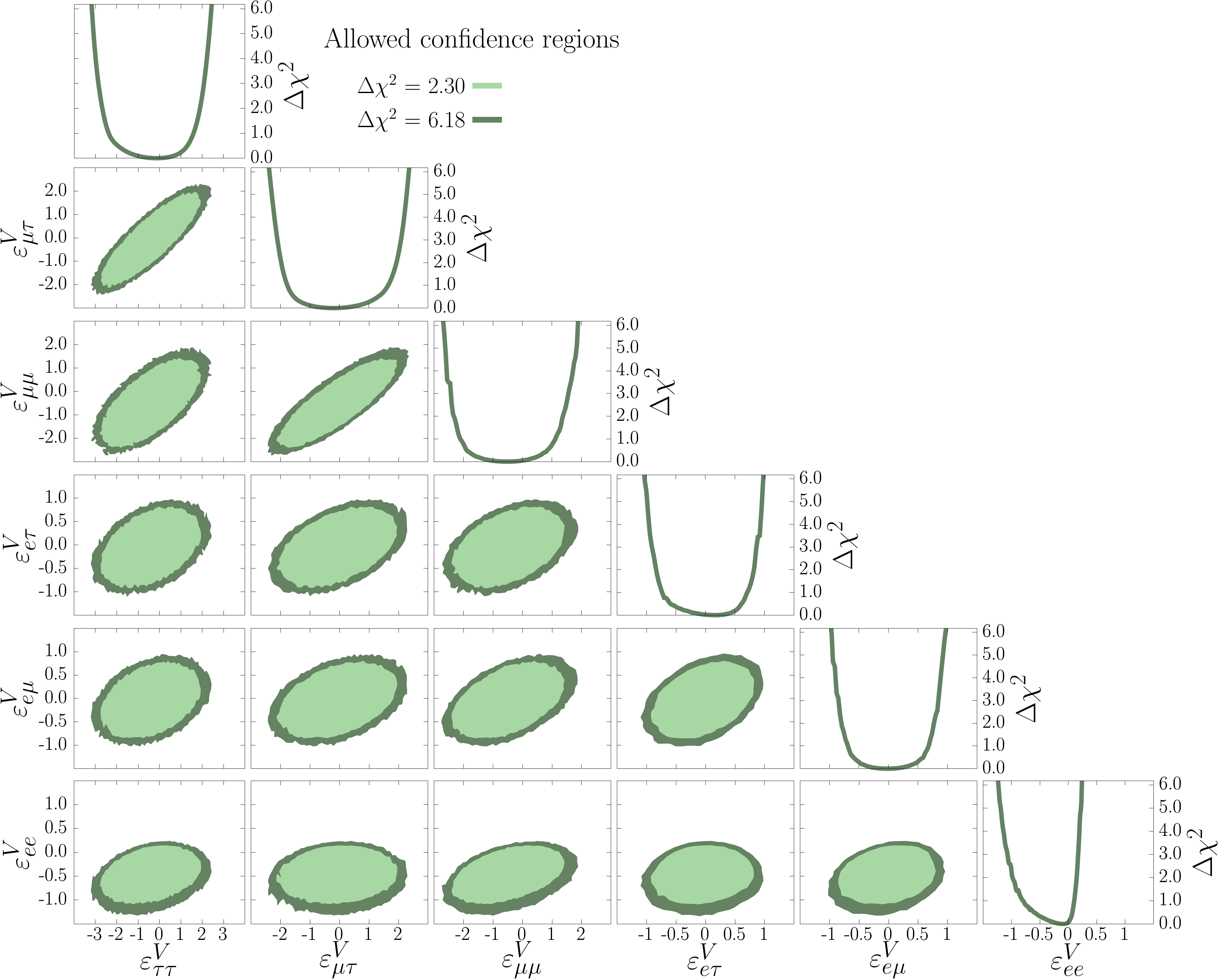}
  \caption{Same as Fig.~\ref{fig:NSIcorner-a}, but for vector NSI with
    electrons ($\Eps_{\alpha\beta}^L = \Eps_{\alpha\beta}^R$).}
  \label{fig:NSIcorner-v}
\end{figure}

Our allowed regions for vector NSI are presented in
Fig.~\ref{fig:NSIcorner-v}, where we see that the potential of
Borexino to test NSI with electrons is again limited in this
multi-parameter scenario: Borexino is only able to constrain NSI at
the level of $|\Eps_{\alpha\alpha}^V| < \mathcal{O}(2-3)$ for
flavour-diagonal interactions, and $|\Eps_{\alpha\beta}^V| <
\mathcal{O}(1)$ for the off-diagonal parameters.  Precise values for
the allowed regions at 90\% CL are given in Tab.~\ref{tab:VA}.

The features describing the allowed regions in the two-dimensional
projections of the $\Delta\chi^2$ are qualitatively similar to the
results found for the axial case, albeit with some differences.  In
particular the regions for $\Eps_{\mu\mu}^V$ and $\Eps_{\tau\tau}^V$
are slightly more symmetric than the axial case.  Again, this can be
qualitatively understood if the cross section in
Eq.~\eqref{eq:nsi-elec} is rewritten in terms of the axial and vector
couplings which results in Eq.~\eqref{eq:nsi-cv-ca} but replacing
$c_{A\alpha} \leftrightarrow c_{V\alpha}$ and $\Eps^A_{\alpha\alpha}
\leftrightarrow \Eps^V_{\alpha\alpha}$.  Thus, a similar interference
as in the axial case takes place here; however since $c_{V\mu} =
c_{V\tau} \sim -0.04$ the quasi-degenerate \emph{SM-like} solution
lies closer to the SM so the region is more symmetric about zero.

For completeness, we also provide our results for NSI involving only
left-handed and right-handed electrons in Tab.~\ref{tab:LR}, where we
again show the results obtained turning on only one operator at a
time, and allowing all NSI operators simultaneously in the fit.  We
have numerically checked that, if only a single operator is included,
we recover the results of the Borexino collaboration for both
left-handed and right-handed NSI (Fig.~5 in
Ref.~\cite{Borexino:2019mhy}) with a very good accuracy. The only
difference is the secondary minimum for $\Eps^L_{ee}\sim -1.3$ which
is outside of the range of parameters explored in
Ref.~\cite{Borexino:2019mhy}.

\begin{table}[t]\centering
  \renewcommand{\arraystretch}{1.2}
  \begin{tabular}{|c || c | c ||c | c ||}
    \hline
    & \multicolumn{4}{|c||}{Allowed regions at 90\% CL $(\Delta\chi^2 = 2.71)$}
    \\
    \cline{2-5} &
    \multicolumn{2}{|c||}{Left-handed} &
    \multicolumn{2}{|c||}{Right-handed}
    \\
    \cline{2-5}
    & 1 Parameter & Marginalized
    & 1 Parameter & Marginalized
    \\
    \hline
    $\Eps_{ee}$
    & $[-1.37, -1.29] \oplus [-0.03, +0.06]$ & $[-1.48, +0.09]$
    & $[-0.23, +0.07]$ & $[-0.53, +0.09]$
    \\
    $\Eps_{\mu\mu}$
    & $[-0.20, +0.13] \oplus [+0.58, +0.81]$ & $[-1.59, +2.21]$
    & $[-0.36, +0.37]$ & $[-1.26, +0.77]$
    \\
    $\Eps_{\tau\tau}$
    & $[-0.26, +0.26] \oplus [+0.45, +0.86]$ & $[-2.00, +2.60]$
    & $[-0.58, +0.47]$ & $[-1.45, +1.04]$
    \\
    $\Eps_{e\mu}$
    & $[-0.17, +0.29]$ & $[-0.78, +0.75]$ & $[-0.21, +0.41]$
    & $[-0.43, +0.42]$
    \\
    $\Eps_{e\tau}$
    & $[-0.26, +0.23]$ & $[-0.81, +0.79]$
    & $[-0.35, +0.31]$ & $[-0.39, +0.43]$
    \\
    $\Eps_{\mu\tau}$
    & $[-0.62, -0.52] \oplus [-0.09, +0.14]$ & $[-1.95, +1.91]$
    & $[-0.26, +0.23]$ & $[-1.10, +1.03]$
    \\
    \hline
  \end{tabular}
  \caption{Constraints at 90\% CL (for 1 d.o.f., 2-sided) on the
    coefficients of NSI operators involving only left-handed
    ($\Eps^L$) or right-handed ($\Eps^R$) electrons. Results are
    provided separately for two cases: when only one NSI operator is
    included at a time (``1 Parameter'') or when the remaining NSI
    coefficients are allowed to float freely in the fit
    (``Marginalized'').}
  \label{tab:LR}
\end{table}

\subsection{Light vector mediators}
\label{sec:light-vec}

The potential of Borexino to probe simplified models with light vector
mediators has been demonstrated in the literature for the $B-L$
case~\cite{Harnik:2012ni, Bilmis:2015lja}, and the bound has been
later recast to other $U(1)'$ models in Refs.~\cite{Kaneta:2016uyt,
  Bauer:2018onh}.  However, Refs.~\cite{Harnik:2012ni, Bilmis:2015lja}
date from 2012 and 2015 and therefore did not include Phase-II data.
Furthermore, their bound was only approximate: it was derived
requiring that the new physics contribution to the event rates should
not exceed the SM expectation.  Here we report our limits for
simplified models, for our data analysis which includes full spectral
information for the Phase II data and a careful implementation of
systematic uncertainties.

\begin{figure}[t]\centering
  \includegraphics[width=0.98\textwidth]{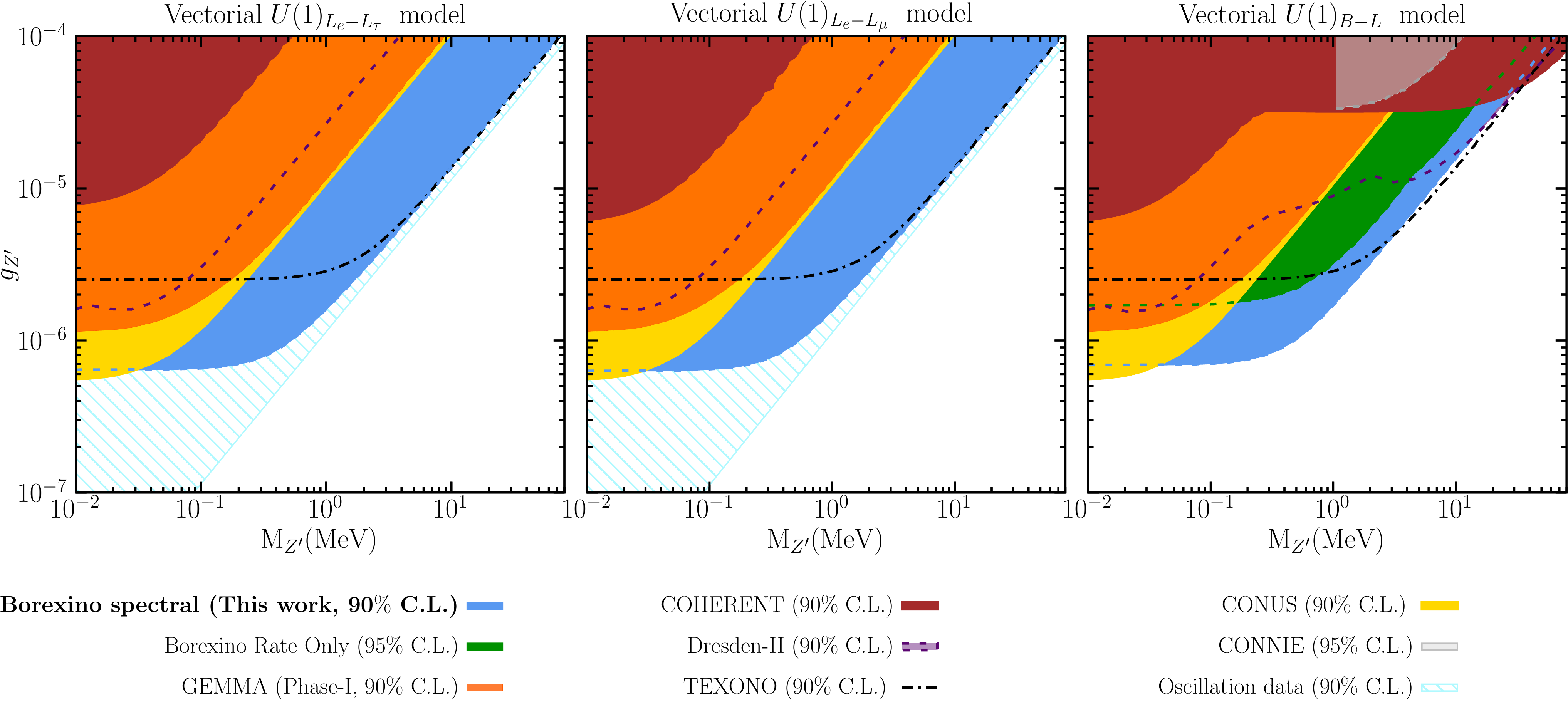}
  \caption{Bounds from our analysis of Borexino Phase II spectral data
    (at 90\% CL for 1 d.o.f., using two-sided intervals,
    \textit{i.e.}, $\Delta\chi^2=2.71$) on the vector mediators
    associated to a new $U(1)'$ symmetry, as indicated in the top
    label for each panel.  Our results are compared to our derived
    bounds from GEMMA~\recite{Beda:2010hk, Coloma:2020gfv},
    TEXONO~\recite{Deniz:2009mu, Coloma:2020gfv},
    COHERENT~\recite{COHERENT:2017ipa, COHERENT:2020iec,
      Coloma:2022avw}, and Dresden-II reactor
    experiment~\recite{Colaresi:2022obx, Coloma:2022avw}, as well as
    those derived from a global fit to oscillation
    data~\recite{Coloma:2020gfv}, all of them obtained with the same
    statistical criterion.  We also show the bounds from
    CONNIE~\recite{CONNIE:2019xid} (95\% CL) and
    CONUS~\recite{CONUS:2021dwh} (90\% CL) which we have obtained
    applying an overall rescaling factor to their published bounds on
    a universally coupled vector (see text for details).  For the
    $B-L$ model, the line labelled ``Borexino Rate Only'' shows the
    approximate bound derived (at 95\% CL, 1 d.o.f., for one-sided
    intervals) in Ref.~\recite{Bilmis:2015lja} (see also
    Ref.~\recite{Harnik:2012ni}).}
  \label{fig:vector-med}
\end{figure}

Our results for vector mediators are presented in
Fig.~\ref{fig:vector-med} for three anomaly-free models in which the
vector couples to $B-L$, $L_e-L_\mu$, or $L_e-L_\tau$.  As seen in the
figure the results for the three models are very similar.  This can be
easily understood from Eqs.~\eqref{eq:ES-prob} and~\eqref{eq:csvece}.
In particular the dominant BSM contribution to the event rates arises
from the interference term with the SM.  So, the new physics will lead
to a modification to the expected number of events that goes as
\begin{equation}
  \Delta N_\text{ev} \propto \sum_\beta
  P_{e\beta} q_{Z'}^{\nu_\beta} \, q_{Z'}^e\, c_{V\beta}
  \sim P_{ee}\, q_{Z'}^{\nu_e}\, q_{Z'}^e\, c_{Ve} \,,
\end{equation}
where we have removed the sum over neutrino flavours, as $c_{Ve}\sim
1\gg c_{V\mu(\tau)}\sim 0.04$ and $P_{ee} > P_{e\mu} \simeq
P_{e\tau}$.  Since for the three models considered $q_{Z'}^{\nu_e}\,
q_{Z'}^e = 1$, the derived bounds are about the same in the three
cases.

For completeness, we also compare our Borexino results to previous
bounds obtained in the literature.  For the range of couplings
considered here, the most relevant constraints come from CE$\nu$NS,
neutrino scattering on electrons, and from neutrino oscillation data,
while measurements of neutrino scattering on nuclei or electrons at
high momentum transfer will be suppressed with the inverse of the
momentum transfer squared (see, \textit{e.g.}, the discussion in
Ref.~\cite{Coloma:2017egw}).  In principle, additional bounds are
obtained from fixed-target experiments or colliders (for example, the
$Z'$ may be produced on-shell and decay to pairs of charged leptons).
However, these bounds are \emph{a priori} model-dependent: for
example, if the new mediator decays predominantly to dark sector
particles, some of them can be significantly relaxed.  Thus, here we
consider only those bounds from neutrino scattering and oscillation
experiments, which always apply to the models considered (for a
detailed compilation of constraints on light vector mediators we refer
the interested reader to, \textit{e.g.}, Refs.~\cite{Bauer:2018onh,
  Coloma:2020gfv}).  Specifically, we consider the following set of
constraints:
\begin{itemize}
\item measurements of $\nu-e$ scattering at the reactor experiments
  GEMMA~\cite{Beda:2010hk} and TEXONO~\cite{Deniz:2009mu}, computed as
  in Ref.~\cite{Coloma:2020gfv}.  Here we note that our analysis of
  GEMMA Phase-II shows a poor fit to the data (with or without BSM
  contributions).  Naively including this dataset in the
  $\Delta\chi^2$ analysis results into strong bounds on the BSM models
  which, however, are not statistically robust.  Thus, the results
  shown here for GEMMA correspond to Phase-I data alone,\footnote{We
  believe that this may explain why our bounds are somewhat more
  conservative than the ones reported in Refs.~\cite{Bilmis:2015lja,
    Harnik:2012ni, Lindner:2018kjo}.} for which (under the SM
  hypothesis) we numerically find $\chi^2/n_\text{d.o.f.} \sim 26/39$,
  $n_\text{d.o.f.}$ being the number of bins.

\item bounds from CE$\nu$NS measurements (on both CsI and Ar nuclei)
  at COHERENT~\cite{COHERENT:2020iec, COHERENT:2017ipa} and at the
  Dresden-II reactor experiment~\cite{Colaresi:2022obx}, computed as
  in Ref.~\cite{Coloma:2022avw} (based on the analysis performed in
  Refs.~\cite{Coloma:2019mbs, Coloma:2020gfv}).  Note that, although
  these experiments were designed to search for CE$\nu$NS (which
  dominates the sensitivity to $B-L$), they are also sensitive to
  $\nu$-e elastic scattering and therefore can be used to set
  constraints on the $L_e-L_\mu$ and $L_e-L_\tau$ models as well (see
  Ref.~\cite{Coloma:2022avw} for details).  For concreteness, here we
  show the results for Dresden-II using the quenching factor labelled
  as ``Fef'' in Ref.~\cite{Coloma:2022avw} (qualitatively similar
  results are obtained for the ``YBe'' quenching factor).

\item constraints derived from the impact of new vector mediators on
  the matter potential in neutrino oscillations.  These were obtained
  from a global fit to oscillation data in Ref.~\cite{Coloma:2020gfv}.
  Note that these constraints only apply to flavoured $U(1)'$ models
  (where the neutrino flavours have different charges under the new
  interaction) and therefore are absent in the $B-L$ case.

\item bounds derived from searches for CE$\nu$NS at
  CONNIE~\cite{CONNIE:2019xid}, and from CE$\nu$NS and ES searches at
  CONUS~\cite{CONUS:2021dwh}.  In both cases the bounds were reported
  for a vector mediator which couples universally to all SM fermions.
  The bounds obtained from ES searches coincide for the $B-L$ and the
  universal models, since $q_{Z'}^{\nu_e} \, q_{Z'}^e=1$ in the two
  cases.  Regarding the bounds from CE$\nu$NS searches, the reported
  limits for the universal model are rescaled for the $B-L$ case,
  assuming an average momentum transfer.\footnote{This approximation
  allows to factor out of the cross section the dependence on the BSM
  model parameters, through a proportionality constant that only
  depends on the weak charges of the nucleus in the SM and the BSM
  cases.  We have checked that this approximation is excellent as we
  are able to correctly reproduce their allowed regions for the
  universal vector scenario.}  For CONNIE we quote the results
  obtained using the Lindhard quenching factor~\cite{Lindhard}.  For
  CONUS, on the other hand, we take the results corresponding to a
  quenching factor\footnote{As we will see this choice has no impact
  on the allowed regions for vector mediators since in this case the
  bound is driven by the ES contribution; however, in the scalar case
  some sensitivity arises from CE$\nu$NS searches as well and
  therefore there is a slight dependence on this choice.} for $k=0.16$
  from Fig.~5 in Ref.~\cite{CONUS:2021dwh}.  Notice that the excluded
  regions derived by CONUS in Ref.~\cite{CONUS:2021dwh} are given as
  90\%~CL (see Ref.~\cite{Rink:2022rsx} for details), while for
  CONNIE~\cite{CONNIE:2019xid} the excluded regions are shown at
  95\%~CL.

\item for the $B-L$ model, we also show previous estimates of the
  Borexino constraint computed in Ref.~\cite{Harnik:2012ni,
    Bilmis:2015lja}.  These results are labelled as ``Rate only''.  As
  outlined above, these are obtained using data available prior to
  2015, requiring that the new physics contribution to the total event
  rates in the energy region corresponding to the \Nuc{7}{Be} line
  does not exceed the SM expectation (within reported uncertainties,
  which at the time were at the level of 8\%).  These limits were
  reported as one-sided constraints, at 95\%~CL with seemingly
  1~d.o.f.
\end{itemize}

As can be seen, our results for Borexino qualitatively agree with the
approximate bounds from Ref.~\cite{Bilmis:2015lja}, but lead to an
improved constraint by a factor $\sim 60\%$ on the upper bound on $Z'$
for very light mediators (and by about 30\% on $g_{Z'}/M_{Z'}$ in the
contact-interaction regime).  Also, as seen in the figure, our
analysis of Borexino Phase-II spectra leads to improvements on the
bounds from scattering experiments in most of the parameter space, for
the three models under consideration.  On the other hand, the bounds
from oscillation data prevail as the strongest constraints on the
$L_e-L_\mu$ and $L_e- L_\tau$ models.

\subsection{Light scalar and pseudoscalar mediators}
\label{sec:light-phi}

\begin{figure}[t]\centering
  \includegraphics[width=0.95\textwidth]{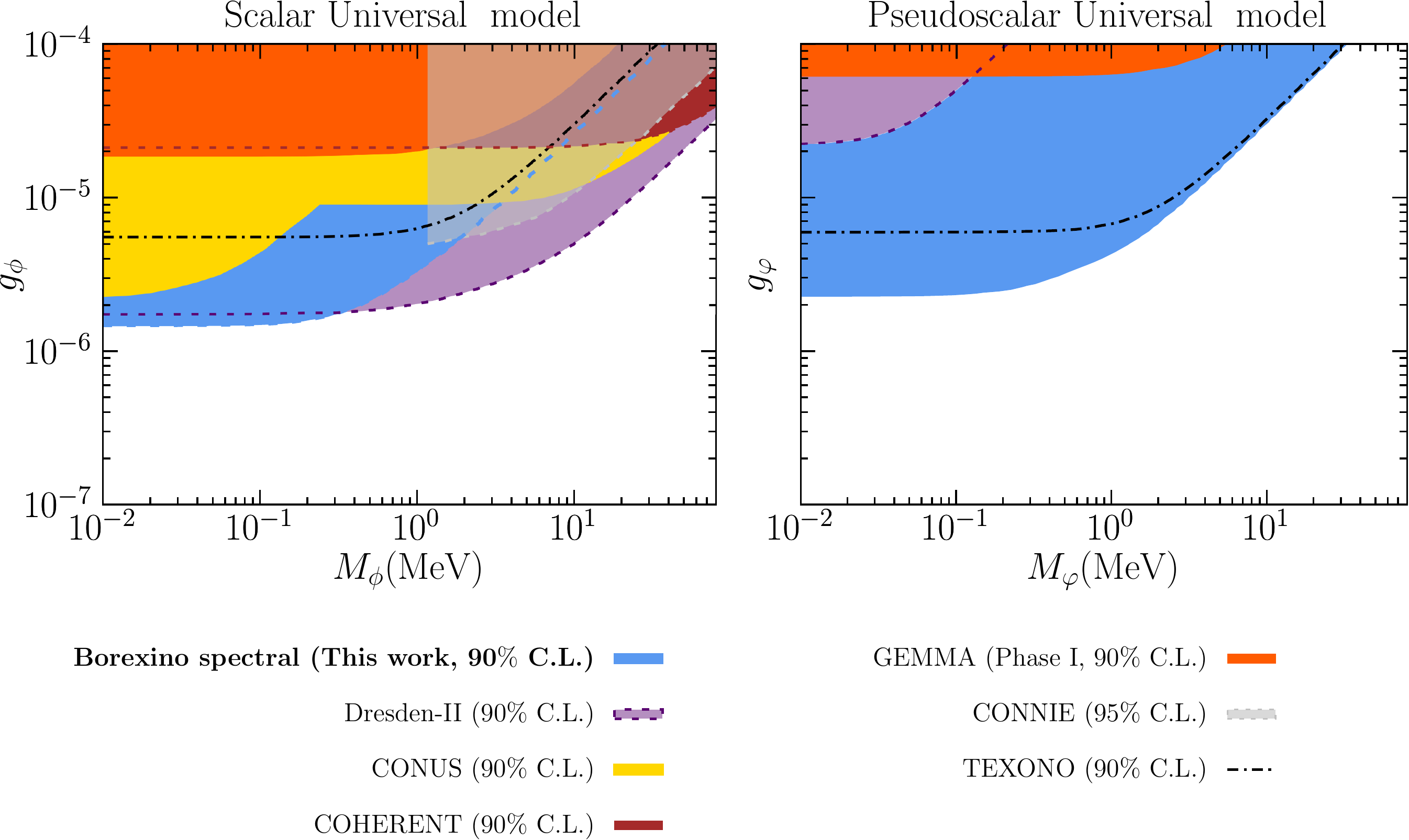}
  \caption{Bounds from our analysis of Borexino Phase-II spectral data
    on a scalar (left panel) and pseudoscalar (right panel) mediator
    (at 90\% CL for 1 d.o.f., using two-sided intervals,
    \textit{i.e.}, $\Delta\chi^2=2.71$) which couples universally to
    all fermions in the SM.  Our results are compared to our derived
    bounds from GEMMA~\recite{Beda:2010hk, Coloma:2020gfv},
    TEXONO~\recite{Deniz:2009mu, Coloma:2020gfv},
    COHERENT~\recite{COHERENT:2017ipa, COHERENT:2020iec,
      Coloma:2022avw}, and Dresden-II reactor
    experiment~\recite{Colaresi:2022obx, Coloma:2022avw}, obtained
    using the same statistical criterion.  For comparison we also show
    the bounds from CONNIE~\recite{CONNIE:2019xid} (95~\%CL) and
    CONUS~\recite{CONUS:2021dwh} (90\%~CL).}
  \label{fig:scalar-med}
\end{figure}

Next let us discuss our results for a scalar (pseudoscalar) mediator
coupled universally to the SM fermions, shown in the left (right)
panel of Fig.~\ref{fig:scalar-med}.  It is important to note that
constraints from oscillation data do not apply to scalar (or
pseudoscalar) mediators.  For comparison, we also show the bounds
derived from neutrino scattering measurements, computed following (or
obtained from) the same references as in Sec.~\ref{sec:light-vec}.
Comparing the scattering constraints shown in
Fig.~\ref{fig:scalar-med} against the same bounds for vector mediators
in Fig.~\ref{fig:vector-med}, we find that, generically, the bounds
imposed on the scalars (and more so for pseudoscalars) are somewhat
weaker.  This is expected because the dominant contribution for the
vector interactions arise from the interference with the SM (which is
quadratic on the new coupling constant) while the scalar (or
pseudoscalar) contributions do not interfere with the SM and therefore
the dependence on the new coupling constant is quartic.  Furthermore
as seen in Eqs.~\eqref{eq:csscale} and~\eqref{eq:cspscale} the cross
section in the scalar and pseudoscalar scenarios are suppressed as
$T_e m_e / E_\nu^2$ and $T_e^2/E_\nu^2$, respectively, with respect to
the vector-mediated scattering cross section in Eq.~\eqref{eq:csvece}.

From the figure we see that our bounds from Borexino Phase II data
improve over previous constraints from ES reactor experiments (GEMMA
and TEXONO) in both panels.  However, for scalar interactions (left
panel) the strongest limits come from the Dresden-II reactor
experiment, and Borexino is only able to give a comparable limit for
very light mediator masses $M_\phi \sim 0.5~\text{MeV}$.  This is so
because for the universal model shown the scalar coupling to quarks
would lead to additional contributions to CE$\nu$NS at the Dresden-II
experiment (as well as for CONNIE and CONUS), which enhances their
sensitivity in the high-mass region (see, \textit{e.g.},
Ref.~\cite{Coloma:2022avw}).  Conversely, in the pseudoscalar scenario
our bounds from Borexino dominate for mediator masses
$M_\varphi\lesssim 5~\text{MeV}$, with TEXONO being the only other
experiment which provides a comparable result.  This is so because
pseudoscalars interactions couple to spin and therefore the
corresponding CE$\nu$NS cross section can be safely neglected.  Thus,
the Dresden-II contour shown in the right panel has been obtained
considering only the new physics contribution to ES (which would pass
their signal selection cuts), leading to a worse constraint compared
to the one obtained in the scalar case.  Finally, note that while in
principle it should also be possible to derive bounds on the
pseudoscalar scenario using ES searches at CONUS, the collaboration
did not study this scenario and therefore no limits from CONUS are
available in this case.  From the results for the scalar case we
expect, however, that these would be weaker than the bound obtained
from Borexino, since the cross section in the pseudoscalar case would
be suppressed to that in the scalar scenario by an additional factor
of $T_e/m_e$ (at CONUS, this factor can be estimated as
$\mathcal{O}(\text{keV}/0.5~\text{MeV}) \sim 2 \times 10^{-3}$).

\section{Conclusions}
\label{sec:conclu}

In this work we have presented a detailed analysis of the spectral
data of Borexino Phase II (corresponding to a total exposure of
$1291.51~\text{days}\times 71.3~\text{tons}$) with the aim at
exploiting its full potential to constrain BSM scenarios.  Following
closely the analysis of the collaboration, we have performed a
detailed treatment of the background subtraction and careful
accounting of all systematic uncertainties (particularly those
associated with the energy reconstruction), including a total of 27
nuisance parameters in our fit.  We have validated our constructed
$\chi^2$ by performing a fit to the solar neutrino fluxes which
accurately reproduces the results of Borexino~\cite{Borexino:2017rsf},
as shown in Figs.~\ref{fig:borexcompa1} and~\ref{fig:borexcompa2}.

We have used our analysis to derive constraints on neutrino magnetic
moments, NSI, and several simplified models with light scalar,
pseudoscalar, and light vector mediators.  In particular, for NSI we
consider at the same time flavour off-diagonal as well as
flavour-diagonal interactions.  This brings the issue of the interplay
between neutrino oscillations and interactions which do not conserve
neutrino flavour, which has been often overlooked in the literature.
We correctly account for those in terms of a density matrix (see
Eq.~\eqref{eq:ES-dens}), which we provide for the specific case of NSI
in Appendix~\ref{sec:appendix}.

Our main results are presented in Sec.~\ref{sec:results}.  For
neutrino magnetic moments we have allowed different magnetic moments
for the different neutrino flavours -- albeit still imposing that they
are flavour-diagonal -- and obtained the bounds in
Eq.~\eqref{eq:mubounds}.  For flavour-independent magnetic moment we
obtained the bound in Eq.~\eqref{eq:mubounds2}.  This is in perfect
quantitative agreement with that obtained by the collaboration in
Ref.~\cite{Borexino:2017fbd} and serves as an additional validation of
our $\chi^2$ implementation.

Our analysis for NSI includes all operators with the same Lorentz
structure: purely vector, axial vector, left-handed only, or
right-handed only (6 parameters in each case, assumed to be real).
Our results are shown in Figs.~\ref{fig:NSIcorner-a}
and~\ref{fig:NSIcorner-v} as well as in Tables~\ref{tab:VA}
and~\ref{tab:LR}.  We have numerically checked that, if only one NSI
operator is included at a time, we recover the results in
Ref.~\cite{Borexino:2017fbd} for the left-handed only and right-handed
only scenarios, to a very good approximation.  However, we find that
when all couplings are allowed simultaneously, strong correlations
appear and the bounds are substantially relaxed, as explicitly shown
in Tables~\ref{tab:VA} and~\ref{tab:LR}.

We have also derived new constraints for simplified models with new
light vector, scalar, and pseudoscalar mediators which we show in
Figs.~\ref{fig:vector-med} and~\ref{fig:scalar-med}.  For mediator
masses $M\lesssim 0.1$ MeV, the obtained experimental bound becomes
independent of the mass of the mediator.  Within this (effectively
massless) regime, the bounds derived from Borexino data for the
considered models read, at 90\% CL (1 d.o.f., one-sided)
\begin{equation}
  \begin{aligned}
    |g_{Z'}^{B-L}| &\leq 6.3\times 10^{-7} \,,
    \\
    |g_{Z'}^{L_e-L_\mu}| &\leq 5.8\times 10^{-7} \,,
    \\
    |g_{Z'}^{L_e-L_\tau}| &\leq 5.8\times 10^{-7} \,,
    \\
    |g_\phi^\text{univ}| &\leq 1.4\times 10^{-6} \,,
    \\
    |g_\varphi^\text{univ}| &\leq 2.2\times 10^{-6} \,.
  \end{aligned}
\end{equation}

Conversely, for sufficiently large mediator masses, the
contact-interaction approximation is recovered and the event rates
vary with a power of $(|g|/M)^n$ with $n=4$ for scalar mediators and
with $n$ ranging between 2 and 4 for vector mediators.  In this limit
we get the following bounds at 90\% CL (1 d.o.f., one-sided)
\begin{equation}
  \begin{aligned}
    |g_{Z'}^{B-L}| \,\big/ M_{Z'}
    & \leq 1.4\times 10^{-6}~\text{MeV}^{-1}
    \quad\text{for}
    & M_{Z'} &\gtrsim\text{few MeV} \,,
    \\
    | g_{Z'}^{L_e-L_\mu}| \,\big/ M_{Z'}
    &\leq 1.2\times 10^{-6}~\text{MeV}^{-1}
    \quad\text{for}
    & M_{Z'} &\gtrsim\text{few MeV} \,,
    \\
    | g_{Z'}^{L_e-L_\tau}| \,\big/ M_{Z'}
    &\leq 1.2\times 10^{-6}~\text{MeV}^{-1}
    \quad\text{for}
    & M_{Z'} &\gtrsim\text{few MeV} \,,
    \\
    | g_\phi^\text{univ}| \,\big/ M_\phi
    &\leq 2.4\times 10^{-6}~\text{MeV}^{-1}
    \quad\text{for}
    & M_\phi &\gtrsim\text{MeV} \,,
    \\
    |g_\varphi^\text{univ}| \,\big/ M_\varphi
    &\leq 2.5\times 10^{-6}~\text{MeV}^{-1}
    \quad\text{for}
    & M_\varphi &\gtrsim\text{MeV} \,.
  \end{aligned}
\end{equation}

\subsection*{Acknowledgements}

This project has received funding/support from the European Union's
Horizon 2020 research and innovation program under the Marie
Skłodowska-Curie grant agreement No 860881-HIDDeN, as well as from
Grants PID2019-105614GB-C21, PID2019-110058GB-C21,
PID2019-108892RB-I00, PID2020-113644GB-I00, ``Unit of Excellence Maria
de Maeztu 2020-2023" award to the ICC-UB CEX2019-000918-M, and Grant
IFT Centro de Excelencia Severo Ochoa No CEX2020-001007-S, funded by
MCIN/AEI/10.13039/501100011033, and by the “Generalitat Valenciana”
grant PROMETEO/2019/087, and by AGAUR (Generalitat de Catalunya) grant
2017-SGR-929.  MCGG is also supported by USA-NSF grant PHY-1915093.
PC is also supported by Grant RYC2018-024240-I funded by MCIN/AEI/
10.13039/501100011033 and by ``ESF Investing in your future''.  SU
acknowledges support from Generalitat Valenciana through the plan GenT
program (CIDEGENT/2018/019).

\appendix

\section{Oscillations with non-standard interactions}
\label{sec:appendix}

In this appendix summarize the quantities relevant for the analysis of
solar neutrino experiments.  We will assume that only the first two
mass eigenstates are dynamical, while the third is taken to be
infinitely heavy.  Since physical quantities have to be independent of the
parameterization of the mixing matrix, we choose a parameterization
that makes analytical expressions particularly simple.  We start from
the Hamiltonian in the flavour basis:
\begin{equation}
  H = U \Delta U^\dagger + V \,,
\end{equation}
where $\Delta = \diag(0, \Dmq_{21}, \Dmq_{31}) / 2E$ and $V = \sqrt{2}
G_F \big[ N_e(x) \diag(1, 0, 0) + \sum_f N_f(x) \Eps_{\alpha\beta}^f
  \big]$.  It is convenient to write $U = O U_{12}$, where $O = R_{23}
R_{13}$ is real while $U_{12}$ is a complex rotation by angle
$\theta_{12}$ and phase $\delta_\text{CP}$.  Then we can write:
\begin{equation}
  H = O \tilde{H} O^\dagger
  \qquad\text{with}\qquad
  \tilde{H} = U_{12} \Delta U_{12}^\dagger
  + O^\dagger V O \,.
\end{equation}
In order to further simplify the analysis, let us now assume that all
the mass-squared differences involving the ``heavy'' states $\nu_3$
can be considered as infinite: $\Dmq_{3l} \to \infty$ for $l=1,2$.  In
leading order, the matrix $\tilde{H}$ takes the effective
block-diagonal form:
\begin{equation}
  \tilde{H} \approx
  \begin{pmatrix}
    H^{(2)} & \boldsymbol{0} \\
    \boldsymbol{0} & \Delta_{33}
  \end{pmatrix}
\end{equation}
where $H^{(2)}$ is the $2\times 2$ sub-matrix of $\tilde{H}$
corresponding to the first and second neutrino states.  Consequently,
the evolution matrix is:
\begin{equation}
  \tilde{S} \approx
  \begin{pmatrix}
    S^{(2)} & \boldsymbol{0} \\
    \boldsymbol{0} & e^{-i \Delta_{33} L}
  \end{pmatrix}
  \quad\text{with}\quad
  S^{(2)} = \Evol\big[H^{(2)}\big]
  \quad\text{and}\quad
  S = O \tilde{S} O^\dagger \,.
\end{equation}
We are interested only in the elements $S_{\alpha e}$.  From the
definition of $O$ we see that $O_{e2} = 0$.  Taking into account the
block-diagonal form of $\tilde{S}$, we obtain:
\begin{equation}
  \label{eq:Smatrix}
  S_{\alpha e} = O_{e1}
  \left( O_{\alpha 1} S_{11}^{(2)}
  + O_{\alpha 2} S_{21}^{(2)} \right)
  + O_{\alpha 3} O_{e3}
  e^{-i\Delta_{33} L} \,.
\end{equation}
In order to derive explicit expression, let us begin writing $H^{(2)}
= H_\text{vac}^\text{(2)} + H_\text{mat}^\text{(2)}$ with:
\begin{align}
  H_\text{vac}^\text{(2)}
  &= \frac{\Dmq_{21}}{4 E_\nu}
  \begin{pmatrix}
    -\cos2\theta_{12} \, \hphantom{e^{-i\delta_\text{CP}}}
    & ~\sin2\theta_{12} \, e^{i\delta_\text{CP}}
    \\
    \hphantom{-}\sin2\theta_{12} \, e^{-i\delta_\text{CP}}
    & ~\cos2\theta_{12} \, \hphantom{e^{i\delta_\text{CP}}}
  \end{pmatrix} ,
  \\
  \label{eq:HmatSol}
  H_\text{mat}^\text{(2)}
  &= \sqrt{2} G_F
  \left[N_e(x)
    \begin{pmatrix}
      c_{13}^2 & 0 \\
      0 & 0
    \end{pmatrix}
    + \sum_f N_f(x)
    \begin{pmatrix}
      -\Eps_D^{f\hphantom{*}} & \Eps_N^f \\
      \hphantom{+} \Eps_N^{f*} & \Eps_D^f
    \end{pmatrix}
    \right],
\end{align}
The coefficients $\Eps_D^f$ and $\Eps_N^f$ are related to the original
parameters $\Eps_{\alpha\beta}^f$ by the relations:
\begin{align}
  \label{eq:eps_D}
  \begin{split}
    \Eps_D^f
    &= c_{13} s_{13}\, \Re\big( s_{23} \, \Eps_{e\mu}^f
    + c_{23} \, \Eps_{e\tau}^f \big)
    - \big( 1 + s_{13}^2 \big)\, c_{23} s_{23}\,
    \Re\big( \Eps_{\mu\tau}^f \big)
    \\
    & \hphantom{={}}
    -\frac{c_{13}^2}{2} \big( \Eps_{ee}^f - \Eps_{\mu\mu}^f \big)
    + \frac{s_{23}^2 - s_{13}^2 c_{23}^2}{2}
    \big( \Eps_{\tau\tau}^f - \Eps_{\mu\mu}^f \big) \,,
  \end{split}
  \\[2mm]
  \label{eq:eps_N}
  \Eps_N^f &=
  c_{13} \big( c_{23} \, \Eps_{e\mu}^f - s_{23} \, \Eps_{e\tau}^f \big)
  + s_{13} \left[
    s_{23}^2 \, \Eps_{\mu\tau}^f - c_{23}^2 \, \Eps_{\mu\tau}^{f*}
    + c_{23} s_{23} \big( \Eps_{\tau\tau}^f - \Eps_{\mu\mu}^f \big)
    \right].
\end{align}
Finally, it is convenient to introduce the following effective
quantities:
\begin{equation}
  P_\text{osc}^{(2)} \equiv |S_{21}^{(2)}|^2 \,,
  \qquad
  P_\text{int}^{(2)} \equiv \Re\big(
  S_{11}^{(2)} S_{21}^{(2)*} \big) \,,
  \qquad
  P_\text{ext}^{(2)} \equiv \Im\big(
  S_{11}^{(2)} S_{21}^{(2)*} \big) \,,
\end{equation}
and to provide the full matrix expression for $O$:
\begin{equation}
  \label{eq:Omatrix}
  O = \begin{pmatrix}
    c_{13} & 0 & s_{13} \\
    -s_{13} s_{23} & c_{23} & c_{13} s_{23} \\
    -s_{13} c_{23} & -s_{23} & c_{13} c_{23}
  \end{pmatrix}.
\end{equation}

\subsection{Neutrino conversion probabilities}

Let us begin from the flavour conversion probabilities, $P_{\alpha e}
= |S_{\alpha e}|^2$.  From Eq.~\eqref{eq:Smatrix} it is
straightforward to derive:
\begin{equation}
  \label{eq:Probs}
  P_{\alpha e} = |O_{e1}|^2
  \left[ |O_{\alpha 1}|^2 |S_{11}^{(2)}|^2
    + |O_{\alpha 2}|^2 |S_{21}^{(2)}|^2
    + 2 \Re\big( O_{\alpha 1} O_{\alpha 2}
    S_{11}^{(2)} S_{21}^{(2)*} \big) \right]
  + |O_{\alpha 3}|^2 |O_{e3}|^2
\end{equation}
where we have used the fact that the term containing a factor
$e^{-i\Delta_{33} L}$ oscillates very fast, and therefore vanish once
the finite energy resolution of the detector is taken into
account.  This expression can be generically written as:
\begin{equation}
  P_{\alpha e} = c_{13}^2 \big[ A_\alpha P_\text{osc}^{(2)}
    + B_\alpha P_\text{int}^{(2)} \big] + D_\alpha
\end{equation}
with
\begin{equation}
  A_\alpha \equiv
  |O_{\alpha 2}|^2 - |O_{\alpha 1}|^2 \,,
  \qquad
  B_\alpha \equiv
  2\, O_{\alpha 1} O_{\alpha 2} \,,
  \qquad
  D_\alpha \equiv
  \sum_{i=\text{all}} |O_{\alpha i}|^2 |O_{ei}|^2 \,.
\end{equation}
From Eq.~\eqref{eq:Omatrix} we get:
\begin{equation}
  \begin{tabular}{c@{\quad}|@{\quad}c@{\qquad}c@{\qquad}c}
    ~ & $A_\alpha$ & $B_\alpha$ & $D_\alpha$
    \\
    \hline
    $\alpha = e$
    & $-c_{13}^2$
    & $0$
    & $1 - (2 c_{13}^2 s_{13}^2)$
    \\[1mm]
    $\alpha = \mu$
    & $c_{23}^2 - s_{13}^2 s_{23}^2$
    & $-2 s_{13} c_{23} s_{23}$
    & $(2 c_{13}^2 s_{13}^2) s_{23}^2$
    \\[1mm]
    $\alpha = \tau$
    & $s_{23}^2 - s_{13}^2 c_{23}^2$
    & $+2 s_{13} c_{23} s_{23}$
    & $(2 c_{13}^2 s_{13}^2) c_{23}^2$
  \end{tabular}
\end{equation}

\subsection{Neutrino density matrix}

Let us now turn to the expression for the density matrix,
$\rho_{\alpha\beta}^{(e)} = S_{\alpha e} S_{\beta e}^*$.  Again, from
Eq.~\eqref{eq:Smatrix} we get:
\begin{multline}
  \rho_{\alpha\beta}^{(e)} = |O_{e1}|^2
  \Big[ O_{\alpha 1} O_{\beta 1} |S_{11}^{(2)}|^2
    + O_{\alpha 2} O_{\beta 2} |S_{21}^{(2)}|^2
    + O_{\alpha 1} O_{\beta 2} S_{11}^{(2)} S_{21}^{(2)*}
    + O_{\alpha 2} O_{\beta 1} S_{11}^{(2)*} S_{21}^{(2)}
    \Big]
  \\
  + O_{\alpha 3} O_{\beta 3} |O_{e3}|^2
\end{multline}
in full analogy with Eq.~\eqref{eq:Probs}.  As before, this expression
can generically be written as:
\begin{equation}
  \rho_{\alpha\beta}^{(e)} = c_{13}^2 \big[
    A_{\alpha\beta} P_\text{osc}^{(2)}
    + B_{\alpha\beta} P_\text{int}^{(2)}
    + i C_{\alpha\beta} P_\text{ext}^{(2)} \big]
  + D_{\alpha\beta}
\end{equation}
with
\begin{equation}
  \begin{aligned}
    A_{\alpha\beta} &\equiv
    O_{\alpha 2} O_{\beta 2} - O_{\alpha 1} O_{\beta 1} \,,
    &\qquad
    B_{\alpha\beta} &\equiv
    O_{\alpha 1} O_{\beta 2} + O_{\alpha 2} O_{\beta 1} \,,
    \\
    C_{\alpha\beta} &\equiv
    O_{\alpha 1} O_{\beta 2} - O_{\alpha 2} O_{\beta 1} \,,
    &\qquad
    D_{\alpha\beta} &\equiv
    \sum_{i=\text{all}} O_{\alpha i} O_{\beta i} |O_{ei}|^2 \,.
  \end{aligned}
\end{equation}
General expressions for the $A_{\alpha\beta}$, $B_{\alpha\beta}$,
$C_{\alpha\beta}$ and $D_{\alpha\beta}$ matrices can be easily
obtained.

\section{Details of the Borexino analysis}
\label{sec:borexdetails}

\subsection{Energy Calibration}
\label{sec:calibration}

As mentioned in the text, we perform the fit to Borexino data in terms
of the observed number of hits $N_h$ as estimator of the recoil energy
of the electron scattered by the incoming neutrinos.  Energy
resolution is included as the Gaussian function in
Eq.~\eqref{eq:dist-diff} which gives the probability that an event
with electron recoil energy $T_e$ yields an observed number of hits
$N_h$ in terms of two parameters, the mean value of $N_h$ for a given
$T_e$, $\bar{N}_h(T_e)$, and the width of the probability distribution
$\sigma_h(T_e)$.  To determine these two parameters as functions of
$T_e$ we proceed in in two steps.

\begin{figure}[t]\centering
  \includegraphics[width=\linewidth]{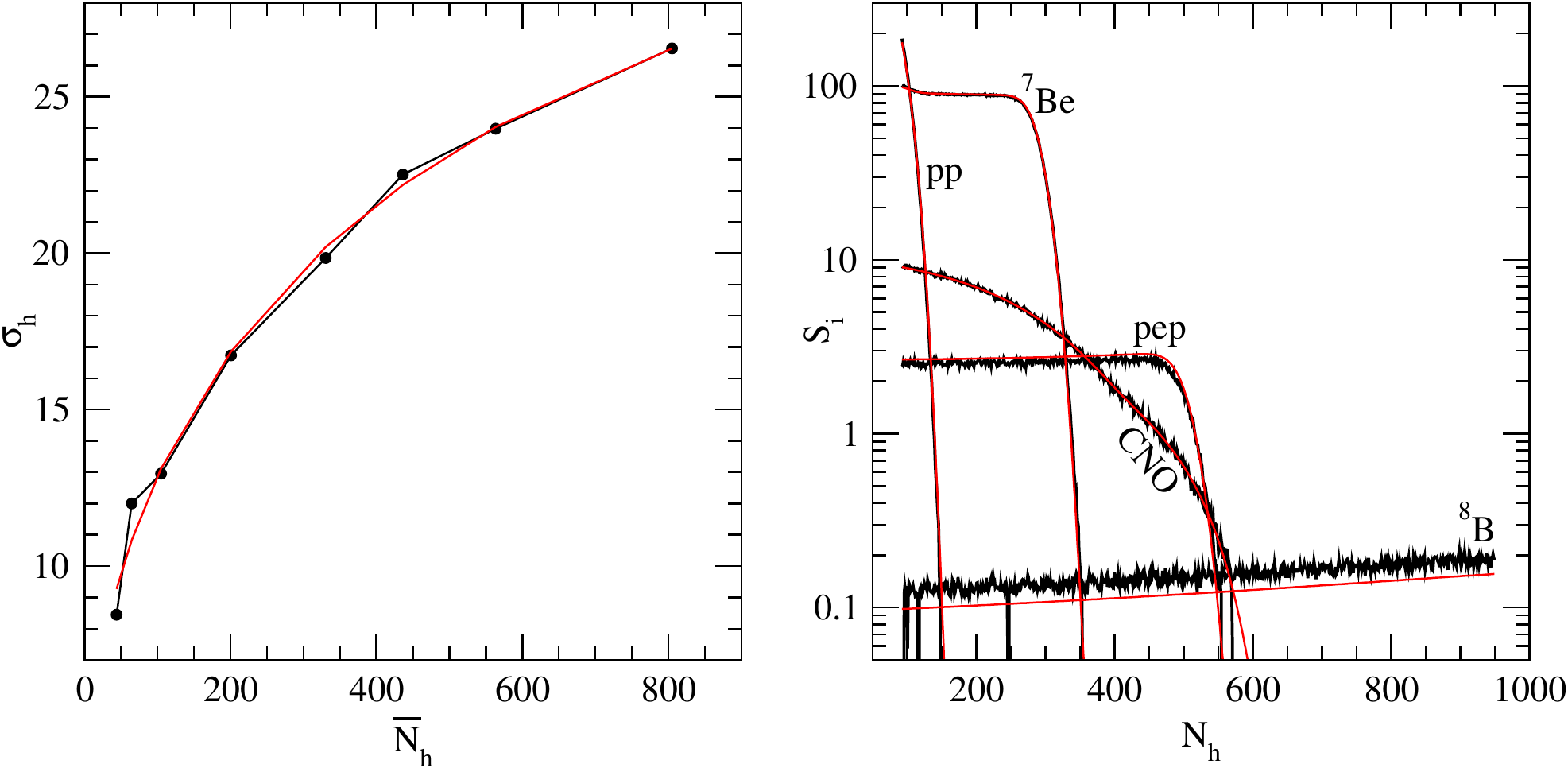}
  \caption{Left: our choice of the energy resolution function, based
    on the relation between $\sigma_h$ and $\bar{N}_h$ inferred from
    the upper panel of Fig.~22 of Ref.~\recite{Borexino:2017rsf}.
    Right: our reconstruction of solar spectra after the optimization
    of the energy scale function.}
  \label{fig:calibra}
\end{figure}

First we obtain the dependence of $\sigma_h\left(\bar{N}_h\right)$
from the calibration results of Borexino published in
Ref.~\cite{Borexino:2017mly}.  Concretely, by fitting the data from
the upper panel of Fig.~22 of the published version (Fig.~21 in the
arXiv version) with Gaussian distributions, we derive a relation
between $\bar{N}_h$ and $\sigma_h$, which is depicted in the left
panel of Fig.~\ref{fig:calibra} and which reads:
\begin{equation}
  \label{eq:sigma-poly}
  \sigma_h(\bar{N}_h)
  = 1.21974 + 1.31394 \sqrt{\bar{N}_h} - 0.0148585 \bar{N}_h \,.
\end{equation}
The calibration results were obtained with a $\gamma$-ray source.  And
the basic assumption is that Eq.~\eqref{eq:sigma-poly} holds equally
for electrons.

Second we introduce that resolution function, which is now a function
of $\bar{N}_h$, in our evaluation of the expected event rates for the
different solar fluxes and infer the dependence of $\bar{N}_h(T_e)$
which best reproduces the solar spectra depicted by Borexino in Fig.~7
of Ref.~\cite{Borexino:2017rsf}.  We assume a 3rd-degree polynomial
relation and find the best agreement to be with
\begin{equation}
  \label{eq:Nh-poly}
  \bar{N}_h(T_e) = -8.065244
  + 493.2560 \bigg[ \frac{T_e}{\text{MeV}} \bigg]
  - 64.09629 \bigg[ \frac{T_e}{\text{MeV}} \bigg]^2
  + 4.146102 \bigg[ \frac{T_e}{\text{MeV}} \bigg]^3 .
\end{equation}
The agreement we find between the Borexino solar spectra and our own
calculations can be seen in the right panel of Fig.~\ref{fig:calibra}.

\subsection{Implementation of systematics uncertainties, and assumed priors}
\label{sec:systematics}

In our analysis we introduce a total of 22 uncorrelated pulls to
include background normalization uncertainties, signal and background
systematic uncertainties related to the detector performance, and
energy shift uncertainties.  In addition 5 correlated pulls are used
to introduce the SSM priors on the solar flux components.  Here we
discuss the implementation these different types of uncertainties
separately.

\subsubsection*{Background normalizations}

We consider 9 different background sources.  Seven correspond to
radioactive isotopes contaminating the scintillator, \Nuc{14}{C},
\Nuc{11}{C}, \Nuc{10}{C}, \Nuc{210}{Po}, \Nuc{210}{Bi}, \Nuc{85}{Kr},
and \Nuc{6}{He}.  Two additional additional backgrounds are due to
pile-up of uncorrelated events, and residual external backgrounds (we
refer in what follows to this last one as ``ext'').

We read the 18 (9 for each data sample) spectra and best fit
normalization for these backgrounds from the results in Fig.~7 of
Ref.~\cite{Borexino:2017rsf}.  Following the procedure described in
that reference we allow their normalization to vary in the fit.  We do
so by introducing a set of 11 independent pulls of which 7
corresponding to the normalizations of \Nuc{14}{C}, \Nuc{10}{C},
\Nuc{210}{Bi}, \Nuc{85}{Kr}, \Nuc{6}{He}, pile-up and ext which are
assumed to be the same for both data samples, and 4 to the
normalization of \Nuc{11}{C}, \Nuc{210}{Po} which are left to be
different in the tagged and the subtracted samples.  The
normalizations corresponding to \Nuc{10}{C}, \Nuc{210}{Bi},
\Nuc{6}{He}, $\Nuc{11}{C}_\text{tag}$, $\Nuc{210}{Po}_\text{tag}$, and
$\Nuc{210}{Po}_\text{sub}$ are left unconstrained and therefore for
them no bias factor is included in the $\chi^2$.  For the other
normalizations we include priors following the information provided in
Ref.~\cite{Borexino:2017rsf}.

The effect of these 11 pulls is included as  described
in Eq.~\eqref{eq:tsys} in terms of the corresponding derivatives
\begin{equation}
  D_{s,i}^c = \sigma_c\, B_{s,i}^c
  \quad\text{for}\quad
  c \in \{ \Nuc{14}{C},
  \Nuc{11}{C}_\text{sub},
  \Nuc{85}{Kr}, \text{pile-up}, \text{ext} \}
\end{equation}
with
\begin{equation}
  \sigma_\text{C14} = 0.05,
  \quad
  \sigma_\text{C11,sub} = 0.0357,
  \quad
  \sigma_\text{Kr} = 0.0368,
  \quad
  \sigma_\text{ext} = 0.0484,
  \quad
  \sigma_\text{pile-up} = 0.02,
\end{equation}
and
\begin{equation}
  D_{s,i}^c = B_{s,i}^c
  \quad\text{for}\quad
  c \in \{ \Nuc{11}{C}_\text{tag}, \Nuc{10}{C},
  \Nuc{210}{Po}_\text{tag}, \Nuc{210}{Po}_\text{sub},
  \Nuc{210}{Bi}, \Nuc{6}{He} \}
\end{equation}
where for convenience we label the $r=1,\dots, 11$ pulls by a label
$c$ which specifies the background component they normalize.

\subsubsection*{Uncorrelated detector related signal and background uncertainties}

We introduce eight pulls to parametrize the systematic uncertainties
related to detector performance and background modelling.  Their
corresponding derivatives are:
\begin{equation}
  \label{eq:deriv-syst}
  D_{s,i}^r = \sum_f \sigma_{f}^r\,S_{s,i}^f
  + \sum_c \sigma_{c}^r \, B_{s,i}^c \, ,
\end{equation}
where $\sigma_f^r$ and $\sigma_c^r$ denote the assumed uncertainties
for their effect each signal ($S_{s,i}^f$) and background
($B_{s,i}^c$) component, respectively.  The values used in our fit are
provided in the upper and lower sections of Tab.~\ref{tab:sys},
respectively.

\begin{table}\centering
  \begin{tabular}{|@{~~}l@{~~}|@{~~}c@{\qquad}c@{\qquad}c@{\qquad}c@{\qquad}c@{\qquad}c@{\qquad}c@{\qquad}c@{~~}|}
    \hline
    & $\xi_1$ & $\xi_2$ & $\xi_3$ & $\xi_4$ & $\xi_5$ & $\xi_6$ & $\xi_7$ & $\xi_8$ 
    \\
    \hline
    pp      & 1.2\% & 2.5\% & 1.5\% & 3.0\% & 2.5\% & 2.2\% & 0.5\% & 0.853\%
    \\
    Be      & 0.2\% & 0.1\% &  ---  & 0.1\% & 0.6\% & 0.2\% & 0.5\% & 0.853\%
    \\
    pep     & 4.0\% & 2.4\% &  ---  & 1.0\% & 4.8\% & 1.6\% & 0.5\% & 0.853\%
    \\
    CNO     & 4.0\% & 2.4\% &  ---  & 1.0\% & 4.8\% & 1.6\% & 0.5\% & 0.853\%
    \\
    B       &  ---  &  ---  &  ---  &  ---  &  ---  &  ---  & 0.5\% & 0.853\%
    \\
    \hline
    C14     &  ---  & 2.5\% &  ---  &  ---  &  ---  &  ---  &  ---  & 0.853\%
    \\
    pile-up &  ---  &  ---  & 1.5\% &  ---  &  ---  &  ---  &  ---  & 0.853\%
    \\
    C11     &  ---  &  ---  &  ---  &  ---  &  ---  &  ---  &  ---  & 0.853\%
    \\
    C10     &  ---  &  ---  &  ---  &  ---  &  ---  &  ---  &  ---  & 0.853\%
    \\
    Po      &  ---  &  ---  &  ---  &  ---  &  ---  &  ---  &  ---  & 0.853\%
    \\
    Bi      &  ---  &  ---  &  ---  &  ---  &  ---  &  ---  &  ---  & 0.853\%
    \\
    Kr      &  ---  &  ---  &  ---  &  ---  &  ---  &  ---  &  ---  & 0.853\%
    \\
    He      &  ---  &  ---  &  ---  &  ---  &  ---  &  ---  &  ---  & 0.853\%
    \\
    Ext     &  ---  &  ---  &  ---  &  ---  &  ---  &  ---  &  ---  & 0.853\%
    \\
    \hline
  \end{tabular}
  \caption{Prior uncertainties assumed for each signal component
    $\sigma_f^r$ (upper part of the table) and for each background
    component $\sigma_{c,\text{sub}}^r = \sigma_{c,\text{tag}}^r$
    (lower part), for the eight nuisance parameters used to account
    for systematic uncertainties in the fit.  Here, $\xi_1 - \xi_6$
    correspond to the six first sources of systematic uncertainties
    listed in Tab.~IV of Ref.~\recite{Borexino:2017rsf} where they
    only show their effect on the determination of the solar fluxes in
    the Borexino fit.  $\xi_2$ corresponds to the choice of energy
    estimator which we find to be also to be relevant for the
    determination of the steeply falling $\Nuc{14}{C}$ background.
    $\xi_3$ corresponds to the pile-up modelling so we assign similar
    uncertainties for the effect of that pull on the pile-up
    background.  $\xi_8$ corresponds to the combined uncertainty on
    the live time, scintillator density and fiducial volume
    uncertainty from the same table which we assume affect all signals
    and backgrounds.  $\xi_7$ refers to an overall normalization error
    of the elastic scattering cross-section~\recite{Bahcall:1995mm}
    (which only affects the signal).}
  \label{tab:sys}
\end{table}

\subsubsection*{Energy shift uncertainties}

In addition to the systematic uncertainties describe above we also
introduce a set of pulls to account for possible systematic shifts in
the energy spectra for the different backgrounds that we have read
from Fig.~7 of Ref.~\cite{Borexino:2017rsf}.  We came out to the need
of this additional pulls by noticing that the Borexino data shown in
that figure can be extracted in two different ways: directly from the
upper panels in that figure, or using the residuals provided in the
lower panels, and assuming that these have been obtained with respect
to the best-fit lines shown in the upper panels.  The results obtained
in these two ways should be equivalent but we found that this was not
the case.

From further inspection of the data obtained in these two ways, it was
clear that they were partially shifted in energy with respect to one
another.  This is most evident for the signal and background
components with steepest energy dependence which happen to occur in
the lowest part of the spectrum where the statistics is higher, and
therefore the effect in the quality of the fit most relevant.  We find
that the introduction of an add-hoc energy shift of about $0.25 N_h$
is able to reduce significantly the discrepancy observed between the
two data sets, reconciling the two results.  Thus, in our fit we use
the data extracted from the upper panels in their Fig.~7, but we
include additional pull terms in order to correct for a possible shift
in energy.  Since such energy shift has the largest impact in the
low-energy part of the spectrum (and, most notably, on the pp solar
neutrinos), we introduce three pulls: one for the \Nuc{14}{C} and
pile-up backgrounds, another one for the \Nuc{11}{C} background, and a
third pull for the \Nuc{210}{Po} background.

Their corresponding derivatives are numerically computed as
\begin{equation}
  D_{s,i}^{h} = \frac{B_{s,i+1}^h - B_{s,i-1}^h}{2}
  \quad\text{for}\quad
  h \in \{ \Nuc{14}{C}+\text{pile-up},
  \Nuc{11}{C}, \Nuc{210}{Po} \} \,,
\end{equation}
for the original binning $i=1,\ldots 858$.  So that for our coarse
binning they are
\begin{equation}
  D_{s,k}^h = \sum_i D_{s,i}^{h}
  \quad\text{for $i$ such that}\quad
  N_h^i \in [N_{h,\text{min}}^k, N_{h,\text{max}}^k] \,,
\end{equation}
for $k=1,\dots,96$.

\subsubsection*{Correlated uncertainties on the solar fluxes}

The goal of our analysis of Borexino data is constraining the BSM
scenarios presented in Sec.~\ref{sec:frameworks}.  For that we assume
the solar fluxes to be as given by the Standard Solar Model (SSM).
Technically this is enforced by introducing five pulls for the
normalization of the five flux components of the solar neutrino signal
subject to the priors of the SSM in both their central values,
uncertainties and correlations.  We introduce those by means of a
covariance matrix \textbf{$\Sigma$} which we obtain from
Ref.~\cite{Vinyoles:2016djt, aldoweb}.  In our calculations, we use
the high-metallicity (HZ) solar model for simplicity.\footnote{It
should be noted that this is the model currently favoured by the CNO
measurement at Borexino~\cite{BOREXINO:2020aww}, albeit with a modest
significance.}  So in this case the corresponding derivatives are
simply:
\begin{equation}
  D_{s,i}^{f} = S_{s,i}^f
  \quad\text{for}\quad
  f \in \{ \text{pp}, \Nuc{7}{Be}, \text{pep},
  \text{CNO}, \Nuc{8}{B} \}
\end{equation}

\subsection{Comparison to the results of the collaboration}

\begin{figure}[t]\centering
  \includegraphics[width=\linewidth]{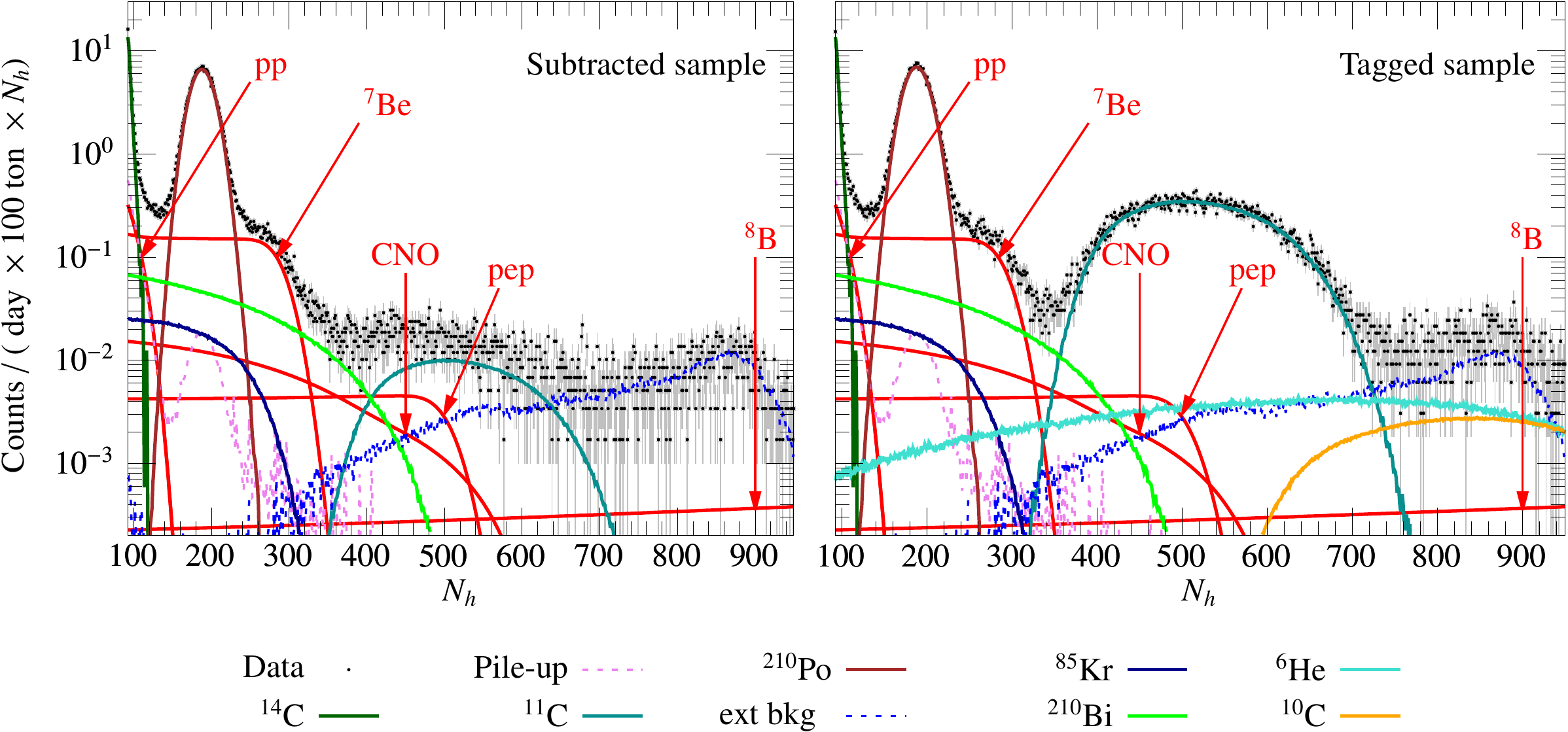}
  \caption{Spectrum for the best-fit normalizations of the different
    components obtained from our fit to the Borexino Phase II data for
    TFC-subtracted (left) and TFC-tagged events.}
  \label{fig:borexcompa1}
\end{figure}

\begin{figure}[t]\centering
  \includegraphics[width=0.85\linewidth]{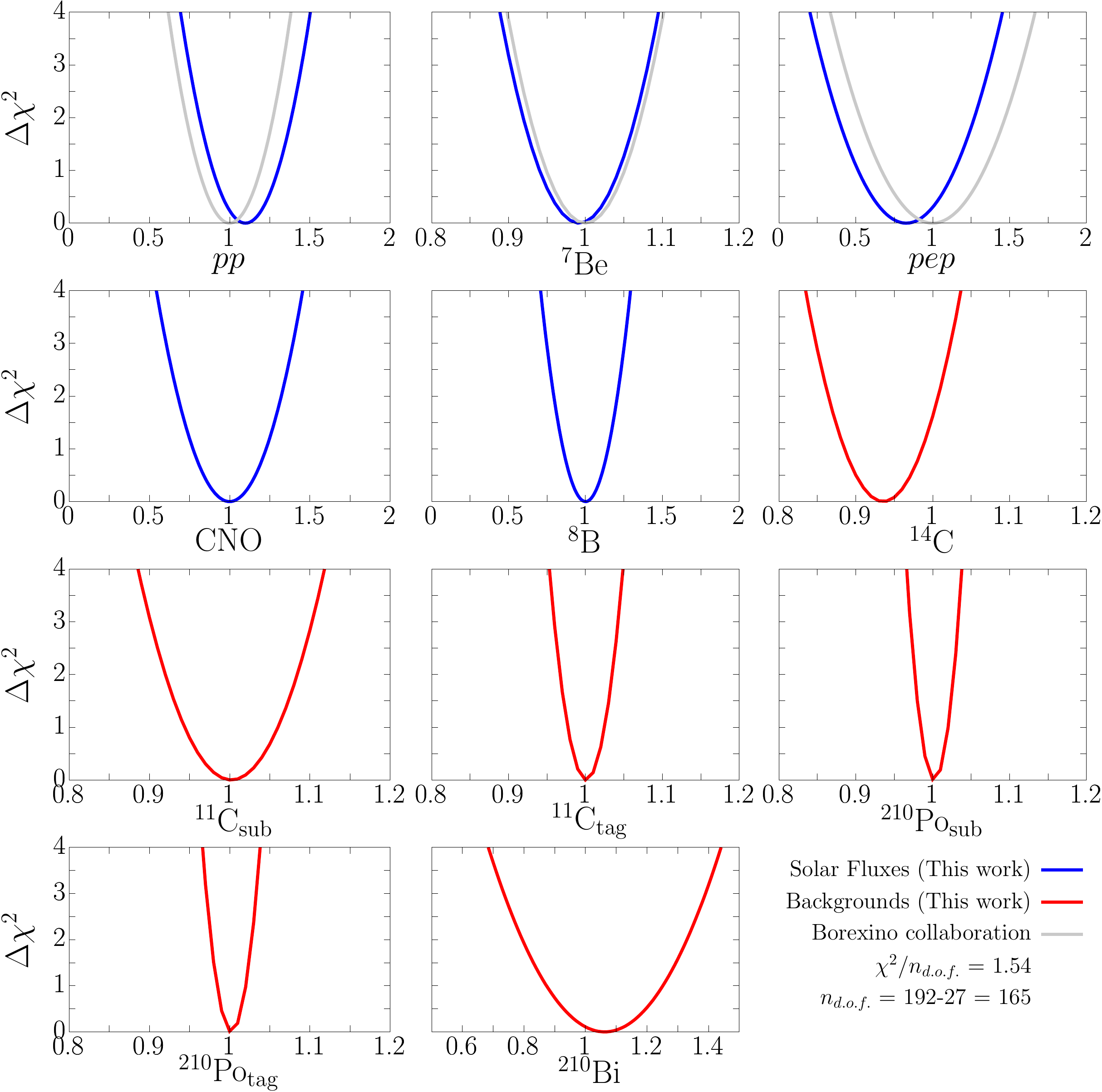}
  \caption{Dependence of $\Delta\chi^2$ of fit to the Borexino
    phase-II spectra on the normalization of the solar fluxes and the
    dominant backgrounds (normalized to the corresponding best fit
    normalizations of the fit of the Borexino collaboration in
    Ref.~\recite{Borexino:2017rsf}).  For comparison we show as light
    blue curves the corresponding results of the determination of
    solar fluxes in Ref.~\recite{Borexino:2017rsf}) (see text for
    details).}
  \label{fig:borexcompa2}
\end{figure}

As a first validation of our constructed $\chi^2$ function we perform
an analysis which intends to reproduce the fit of the collaboration in
Ref.~\cite{Borexino:2017rsf} to determine the solar neutrino fluxes.
So we make a fit in which we do not include the covariance matrix for
the solar fluxes previously described.  Instead we leave the solar pp,
\Nuc{7}{Be} and pep neutrino signal fluxes completely free.  We do
introduce two priors for the CNO and \Nuc{8}{B} the SSM (for
concreteness we chose the high metallicity version of the model) as
Borexino does.  In the analysis we fit all background normalizations
with the priors described above, and include the eight pulls for the
uncorrelated systematics and the the three pulls for the energy shift.
The result for our best fitted spectra (this is, our version of Fig.~7
in Ref.~\cite{Borexino:2017rsf}) is shown in
Fig.~\ref{fig:borexcompa1}.  The dependence of our $\Delta\chi^2$ on
the normalization of the solar fluxes and the dominant backgrounds is
shown in Fig.~\ref{fig:borexcompa2}.  The flux normalizations in the
figure are normalized to the best fit values obtained by Borexino
collaboration, hence a value of ``1'' means perfect agreement between
our best fit fluxes and those of the collaboration.  For comparison we
also show for the solar fluxes as grey lines the inferred
$\Delta\chi^2$ from the results quoted by the collaboration in Table
II of Ref.~\cite{Borexino:2017rsf} (assuming it is approximately
Gaussian).  The figures show that our constructed event rates, the
best-fit normalization of all signal and background components, and
the precision in the determination of the solar fluxes reproduce with
very good accuracy those of the fit performed by the collaboration.
Ref.~\cite{Borexino:2017rsf} does not give enough information for a
quantitative comparison of the $\Delta\chi^2$ for background
normalizations, but we find that the precision we obtain for those is
in good qualitative agreement with the results in Table III and Fig.~5
of Ref.~\cite{Borexino:2017rsf}.

\bibliographystyle{JHEP}
\bibliography{references}

\end{document}